\documentclass{aa}
\usepackage{txfonts}
\usepackage{epsfig}
\usepackage{subfigure}
\usepackage{multirow}

\newcommand{\gapprox}{\hbox{\lower .8ex\hbox{$\,\buildrel > \over\sim\,$}}}
\newcommand{\lapprox}{\hbox{\lower .8ex\hbox{$\,\buildrel < \over\sim\,$}}}
\usepackage{times}
\usepackage{natbib}
\usepackage{rotating}
\bibpunct{(}{)}{;}{a}{}{,}

\begin{document}

\title{Self-similarity of temperature profiles in distant galaxy clusters: the quest for a universal law}

\author{A. Baldi\inst{1,2} 
\and S. Ettori\inst{2,3} 
\and S. Molendi\inst{4}
\and F. Gastaldello\inst{4,5}
}

\offprints{A. Baldi, \email{alessandro.baldi@oabo.inaf.it}}

\institute{Dipartimento di Astronomia, Universit\`a di Bologna, via Ranzani 1, I--40127, Bologna, Italy 
\and INAF, Osservatorio Astronomico di Bologna, via Ranzani 1, I--40127, Bologna, Italy
\and INFN, Sezione di Bologna, viale Berti Pichat 6/2, I--40127 Bologna, Italy
\and INAF-IASF, via Bassini 15, I--20133, Milan, Italy
\and Department of Physics and Astronomy, University of California at Irvine, 4129 Frederick Reines Hall, Irvine, CA 92697--4575, U.S.A.
}

\date{Received 2012}

\titlerunning{Self-similarity of temperature profiles in distant galaxy clusters}
\authorrunning{A. Baldi et al.}

\abstract {We present the XMM-Newton temperature profiles of 12 bright ($L_X>4\times10^{44}$ erg s$^{-1}$) clusters of galaxies at 
$0.4<z<0.9$, having an average temperature in the range $5\la kT\la 11$~keV.} 
{The main goal of this paper is to study for the first time the temperature profiles of a sample of high-redshift
clusters, to investigate their properties, and to define a universal law to describe the temperature radial profiles in galaxy clusters as
a function of both cosmic time and their state of relaxation.} 
{We performed a spatially resolved spectral analysis, using Cash statistics, to measure the temperature in the intracluster 
medium at different radii.} 
{We extracted temperature profiles for the clusters in our sample, finding that all 
profiles are declining toward larger radii. The normalized temperature profiles (normalized by the mean temperature $T_{500}$) 
are found to be generally self-similar. The sample was subdivided into five cool-core (CC) and seven non cool-core (NCC) clusters by introducing a 
pseudo-entropy ratio $\sigma=(T_{IN}/T_{OUT})\times(EM_{IN}/EM_{OUT})^{-1/3}$ and defining the objects with $\sigma<0.6$ as CC clusters and 
those with $\sigma\ge0.6$
as NCC clusters. The profiles of CC and NCC clusters differ mainly in the central regions, with the latter exhibiting a slightly flatter 
central profile. A significant dependence of the temperature profiles on the pseudo-entropy ratio $\sigma$ 
is detected by fitting a function of $r$ and $\sigma$, showing an indication that the outer part of the profiles becomes steeper 
for higher values of $\sigma$ (i.e. transitioning toward the NCC clusters).
No significant evidence of redshift evolution could be found within the redshift range sampled by our clusters ($0.4<z<0.9$). 
A comparison of our high-z sample with intermediate clusters at $0.1<z<0.3$ showed how the CC and NCC cluster temperature profiles
have experienced some sort of evolution. This can happen because higher $z$ clusters are at a less advanced stage of their formation 
and did not have enough time to create a relaxed structure, which is characterized by a central temperature dip in CC clusters and by flatter
profiles in NCC clusters.}
{This is the first time that a systematic study of the temperature profiles of galaxy clusters at $z>0.4$ has been attempted. We 
were able to define the closest possible relation to a universal law for the temperature profiles of galaxy clusters at $0.1<z<0.9$, showing
a dependence on both the relaxation state of the clusters and the redshift.
}


\maketitle

\section{Introduction}

Clusters of galaxies are the largest virialized structures in the
Universe, arising from the gravitational collapse of high peaks of
primordial density perturbations. They represent unique signposts
where the physical properties of the cosmic diffuse baryons can be
studied in great detail and can be used to trace the past history of cosmic
structure formation \citep[e.g.][]{peebles93, coles95, peacock99, 
rosati02, voit05}.
As a result of adiabatic compression and shocks
generated by supersonic motion during shell crossing and
virialization, a hot thin gas permeating the cluster's gravitational
potential well is formed. 
This gas reaches temperatures of several $10^7$ K and therefore emits mainly via
thermal bremsstrahlung and is easily detectable in the X-rays.
The gas key observable quantities are its density, temperature, and metallicity. 
Assuming hydrostatic equilibrium, the gas temperature and density profiles 
allow one to derive the total cluster mass and thus to use galaxy clusters 
as cosmological probes \citep[e.g.][]{voit05}. Temperature and density profiles 
can also be combined to determine the intracluster medium (ICM) entropy distribution, which provides 
valuable information on the cluster thermodynamic history and has proven 
to be a powerful tool in investigating non-gravitational processes \cite[e.g.][]{pratt06}.
Radial temperature profiles of the hot ICM in galaxy clusters are therefore
very important for studying the gravitational processes responsible for large-scale structure formation 
and non-gravitational energy input into the ICM. 

Early measurements of cluster temperature profiles were obtained by ASCA and Beppo-SAX
\citep[e.g.][]{ikebe97,markevitch98,ettori00,finoguenov01,nevalainen01,degrandi02}.
In particular, using ASCA data, \citet{markevitch98} obtained 
temperature profiles for a sample of 32 nearby clusters, which showed significant declines with radius 
between $r=0.1r_{vir}$ and $r=0.6r_{vir}$. These authors also found that in clusters without obvious 
mergers, the radial temperature profiles outside the cool cores were similar 
when normalized to the virial radius. Consistent results were obtained using the spatially resolved
spectroscopic capabilities of BeppoSAX.
In their analysis of 21 clusters, \citet{degrandi02} found declining temperature profiles, 
in good agreement with \citet{markevitch98} outside the cores [$r > 0.15-0.2r_{vir}$], although 
they were less peaked at the center. De Grandi and Molendi also found that for $r>0.2r_{180}$, where the gas can be 
treated as a polytrope, the polytropic index derived for cool-core (CC) clusters is significantly flatter
than the index derived for non cool-core (NCC) clusters, corresponding to $1.20\pm0.06$ for CC clusters and 
to $1.46\pm0.06$ for NCC clusters.

More recently, Chandra and {\em XMM-Newton} allowed measuring the temperature profiles of galaxy clusters
with better accuracy and improved spatial resolution, especially in their central regions
\citep[e.g.][]{vikhlinin05,vikhlinin06b,baldi07,pratt07,leccardi08a}, without the complications 
introduced by the point spread function (PSF) of ASCA and BeppoSAX.
In particular, using Chandra data, \citet{baldi07} were able to study the temperature profiles of 12 clusters in
the redshift range $0.1<z<0.3$, dividing the clusters into CC and NCC clusters 
in a systematic fashion. These authors found that the profiles in the inner $0.1r_{180}$ were showing a positive
gradient $kT(r)\propto r^\mu$, with $\mu\sim0.25$ in CC clusters and $\mu\sim0$ in NCC clusters.
Moreover Baldi et al. found that the outer region profiles were significantly steeper in the NCC systems than in the CC 
systems. This behavior agrees with the recent findings of \citet{arnaud10}, who derived a 
universal pressure profile of galaxy clusters (including CC and NCC clusters) 
in the REXCESS local sample ($z<0.2$) with low dispersion, especially
in the external regions. Steeper temperature gradients in the profiles of NCC systems than in those of CC systems are
indeed expected if this pressure profile, analogous for both cluster categories, is combined with the 
density profiles \citep[steeper in CC and flatter in NCC clusters, see e.g.][]{cavaliere09}.
Probing a similar redshift range, \citet{leccardi08a} analyzed a sample of $\approx$50 galaxy clusters
observed by {\em XMM-Newton} to measure their radial temperature profiles. 
In agreement with previous results, these authors found a decline of the temperature in the $0.2-0.6r_{180}$ range. 
In contrast with \citet{degrandi02} and \citet{baldi07}, Leccardi and Molendi did not find any evidence 
of a dependence in the slope of the outer 
regions from the presence or absence of a cool core, nor did they 
find evidence of profile evolution with the redshift out to $z\approx0.3$.

Cosmological simulations have also tackled the measure of temperature profiles in galaxy clusters.
However, the existing simulations \citep[e.g.][]{loken02,roncarelli06} are not able to account for the physics at small
radii (i.e. in the cluster core regions) and
are not considering the evolution of the profiles through cosmic time.

In this work we aim at studying for the first time the temperature profiles in a sample of galaxy clusters at $z>0.4$, i.e. 
in a redshift range where no comprehensive study is present in the literature.
The plan of the paper is as follows. 
In \S~2 we describe how we selected our {\em XMM-Newton} sample and the data
reduction procedure we followed. 
We describe our spectral analysis strategy in \S~3, with particular
focus on the background treatment procedure (\S~3.1) and the correction
we applied to take into account the PSF of {\em XMM-Newton} (\S~3.2).
In \S~4 we present the results obtained in our analysis, presenting the temperature
profiles of individual clusters in the sample and the classification method we used
to divide them into CC and NCC (\S~4.1), their profiles by the overdensity radius $r_{500}$ and
by the average temperature $T_{500}$, and the properties of self-similarity they exhibit (\S~4.2). 
We also present several fits of the normalized profiles
of the total sample (\S~4.2.1), of the CC and NCC clusters samples individually (\S~4.2.2), 
and of the external regions of the profiles (\S~4.2.3). In \S~5, we discuss our results
and compare our sample with the \citet{leccardi08a} intermediate redshift cluster sample
(\S~5.1) to attempt to define a universal law for temperature profiles for
galaxy clusters at $0.1<z<0.9$ (\S~5.2).
Our conclusions are summarized in \S~6. 

We adopt a cosmological model with $H_0=70$ km/s/Mpc,
$\Omega_m=0.3$, and $\Omega_\Lambda=0.7$ throughout the paper. Confidence
intervals are quoted at $1\sigma$ unless otherwise stated.


\begin{figure}
\includegraphics[width=9.2cm, angle=0]{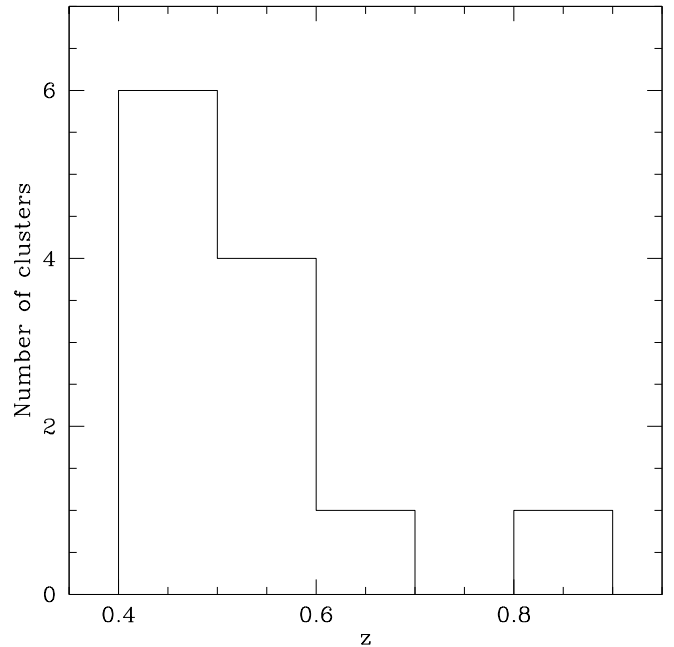}
\caption{Redshift distribution of the {\em XMM-Newton} galaxy cluster bright sample presented in this paper.}
\label{histoz}
\end{figure}

\begin{table*}
\caption{{\em XMM-Newton} archival observations of the 12 bright galaxy clusters analyzed in this paper.\label{exposures}}
\centering
\begin{tabular}{lcrrccccc}
\hline
Cluster & $z$ & Obs. Date & Obs. ID & $t_{clean,MOS1}$ & $t_{clean,MOS2}$ & $t_{clean,pn}$ & $N_H$ & No. of radial bins\\
 & & & & (ksec) & (ksec) & (ksec) & (10$^{20}$ cm$^{-2}$) & considered \\
\hline\hline
\object{A851}              &  0.407   & 2000 Nov 06 & 0106460101 & 41.7 & 41.1 & 30.3 & 1.0 & 7 \\
\hline
\object{RXCJ0856.1+3756}   &  0.411   & 2005 Oct 10 & 0302581801 & 24.2 & 23.9 & 14.7 & 3.2 & 4 \\
\hline
\object{RXJ2228.6+2037}    &  0.412   & 2003 Nov 18 & 0147890101 & 24.6 & 24.4 & 19.3 & 4.3 & 9 \\
\hline
\object{RXCJ1206.2-0848}   &  0.440   & 2007 Dec 09 & 0502430401 & 29.1 & 28.8 & 20.6 & 3.7 & 8 \\
\hline
\object{IRAS09104+4109}    &  0.442   & 2003 Apr 27 & 0147671001 & 12.1 & 12.3 & 8.3 & 1.4 & 4 \\
\hline
\object{RXJ1347.5-1145}    &  0.451   & 2002 Jul 31 & 0112960101 & 32.6 & 32.4 & 27.5 & 4.9 & 9 \\
\hline
\object{CLJ0030+2618}      &  0.500   & 2005 Jul 06 & 0302581101 & 15.0 & 14.0 & - & 3.7 & 3 \\
                  &          & 2006 Jul 27 & 0402750201 & 27.0 & 27.4 & 19.6 &  &  \\
                  &          & 2006 Dec 19 & 0402750601 & 28.5 & 28.6 & 20.6 &  &  \\
\hline
\object{MS0015.9+1609}     &  0.541   & 2000 Dec 29 & 0111000101 & 30.6 & 30.1 & 20.4 & 4.0 & 7 \\
                  &          & 2000 Dec 30 & 0111000201 &  5.5 &  5.4 &  - &  &  \\
\hline
\object{MS0451.6-0305}     &  0.550   & 2004 Sep 16 & 0205670101 & 25.5 & 26.1 & 19.4 & 3.9 & 5 \\
\hline
\object{MACSJ0647.7+7015}  &  0.591   & 2008 Oct 09 & 0551850401 & 52.9 & 53.7 & 32.3 & 5.4 & 6 \\
                  &          & 2009 Mar 04 & 0551851301 & 33.2 & 34.6 & 18.4 &  &  \\
\hline
\object{MACSJ0744.9+3927}  &  0.698   & 2008 Oct 17 & 0551850101 & 40.4 & 41.4 & 22.4 & 5.7 & 6 \\
                  &          & 2009 Mar 21 & 0551851201 & 63.8 & 67.0 & 35.8 &  &  \\
\hline
\object{CLJ1226.9+3332}    &  0.890   & 2001 Jun 18 & 0070340501 & 10.3 & 10.7 &  5.3 & 1.8 & 4 \\
                  &          & 2004 Jun 02 & 0200340101 & 67.2 & 67.5 & 53.4 &  &  \\
\hline
\end{tabular}
\end{table*}

\section{Sample selection and data reduction}\label{sampsel}

In Table~\ref{exposures}, we present the list of galaxy clusters
analyzed in this paper. The clusters considered are picked from the
{\em XMM-Newton} archive sample presented in \citet[][hereafter BAL12]{baldi12}. Here we selected all clusters
with at least 3,000 {\tt MOS} net counts to obtain temperature radial profiles
with a sizeable number of bins (from 3 to 9, depending on the cluster luminosity) 
and reasonable errors on $kT$ ($\sigma_{kT}/kT<15\%$). Thus, a sample of twelve galaxy
clusters was selected, with a redshift range of $0.4\la z \la0.9$
although half of the clusters are located at $z<0.5$ (Figure~\ref{histoz}).
Since this sample was simply taken from the archive without any selection criteria apart from the number of counts available and the
redshift, we caution that our cluster sample is far from being a complete sample and the relative proportion of CC and NCC clusters 
may not be representative of the relative fraction of CC and NCC clusters present in the Universe, within the redshift range explored.
The full details of data reduction are described in BAL12; here we summarize only the main steps
used in processing the data.

The observation data files (ODF) were processed to produce calibrated event 
files using the {\em XMM-Newton} science analysis system ({\tt SAS v11.0.1}). 
As described in BAL12, the {\tt pn} data were not used in the spectral analysis
because of problems in the characterization of the background in {\tt pn} spectra of extended sources 
\citep[see e.g.][and the {\em XMM-ESAS} cookbook\footnote{ftp://xmm.esac.esa.int/pub/xmm-esas/xmm-esas.pdf}]{leccardi08a} and
inconsistencies with the measures of temperature and abundance with the {\tt MOS} detectors.

The soft proton cleaning was performed using a double filtering process. First, we extracted a light curve in 100s bins in 
the 10-12 keV energy band by excluding the central CCD, applied a threshold of 0.20 cts s$^{-1}$, produced a GTI file and 
generated the filtered event file accordingly, to remove the majority of the flares. 
The remaining softer flares (that do not contribute significantly to the emission at $E>10$ keV) were removed by
extracting a light curve in the 2-5 keV. The resulting histogram was fit with a Gaussian distribution with a 
mean count rate $\mu$, and the standard deviation $\sigma$. A 3$\sigma$ clipping algorithm (excluding all time intervals
with a count rate higher than $\mu+3\sigma$) and visual inspection of the light curves were then used 
to remove the remaining background flaring periods.
In Table \ref{exposures} we list the resulting clean exposure times for the {\tt MOS} and {\tt pn} detectors.

After additional filtering was applied to remove events with low spectral quality, we merged the event files from each 
detector to create a single event file and created a 0.5-8 keV band image of each cluster.
The resulting image was used to define the extraction regions for the temperature profiles and to create a list of point sources
(using both the {\tt EBOXDETECT} task and a visual inspection of the field to check for false detections). 
If a point source is located inside an extraction region, 
a circular region of radius $\ga15''$ (depending on the source brightness), centered at the position of the point source
is excised from the spectral extraction region.
As in BAL12, the {\tt pn} data were used to create the merged image to detect the point sources that needed to be removed from
the cluster spectra at a lower flux limit.

\section{Spectral analysis strategy}\label{specan}

The emission from each cluster was subdivided into several concentric annular regions, allowing each annulus to have
a total of at least 1,000 net {\tt MOS} counts and a minimum thickness of $15^{\prime\prime}$. 
The number of annuli considered for each cluster is reported in Table~\ref{exposures}.

\subsection{Background spectral modeling}

The background in our {\tt MOS} spectra was modeled (instead of subtracted) following the procedure developed
by \citet{leccardi08a}, which is also described extensively in BAL12. This procedure is preferrable to a direct 
background subtraction because it allows the use of Cash statistics \citep{cash79} and avoids problems caused by the vignetting 
of the background spectra, which would need to be extracted at larger off-axis angle than the cluster spectra.
Here we summarize just the main aspects of this procedure, more details can be found in \citet{leccardi08a} and in
BAL12.

The background model consists mainly of five components: a thermal emission from the Galaxy halo (HALO), 
the cosmic X-ray background (CXB), a residual from the filtering of quiescent soft protons (QSP), 
the cosmic ray induced continuum (NXB) and the fluorescence emission lines.
The parameters for this model were first estimated (using Cash statistics and considering the 0.7-8 keV band) 
from a spectrum extracted in an annulus located between 
$10^\prime$ and 12$^\prime$ from the center of the field of view for {\tt MOS1} and {\tt MOS2}.
To fit the cluster spectra, the background parameters (and their 1$\sigma$ errors) were then rescaled by the area 
in which the cluster spectra were extracted. Appropriate correction factors 
\citep[][dependent on the off-axis angle]{leccardi08a} were considered for the HALO and CXB components 
and a vignetting factor 
(corresponding to $1.858-0.078r$, with $r$ the distance from the center of the annulus) was applied to the QSP 
component. All these rescaled values (and their 1$\sigma$ errors) were transferred to an XSPEC model with the same background 
components used 
in the fit of the 10$^\prime$-12$^\prime$ annulus plus a thermal {\tt mekal} model \citep{liedahl95} for the emission of the cluster, leaving its 
temperature, abundance and normalization free to vary. The 1$\sigma$ errors on the parameters 
were used to fix a range where the normalizations of the background components are allowed to vary.
This model was tailored to each cluster and used to fit all spectra in our sample. A joint {\tt MOS1} plus {\tt MOS2} fit (using Cash statistics
and therefore requiring simply minimal grouping to avoid spectral bins with no counts) was performed to increase statistics. 
We were able to do that because there are no calibration problems between the two detectors and an independent analysis of 
the spectra from {\tt MOS1} and {\tt MOS2} led to consistent results in the measurements of $kT$ and $Z$.

\subsection{{\em XMM-Newton} PSF correction and spectral fitting}\label{psf}
In the present analysis we also considered the effect that PSF of EPIC {\em XMM-Newton} ($HEW\sim15^{\prime\prime}$)
has on the determination of the temperature profiles in our cluster sample.
This is a different approach than in BAL12, where we were interested only in the measure of the abundances in three different radial
bins ($r<0.15r_{500}$, $0.15r_{500}<r<0.4r_{500}$, $r>0.4r_{500}$), a measurement that we found not to be affected by the PSF.
In the present sample we used the procedure developed in {\em XMM-ESAS} \citep{snowden04} that considers 
the crosstalk ancillary response files (ARFs) between two regions to remove the contribution of each region from the other
region and vice versa. Therefore, for each spectral annulus, we computed the crosstalk ARF relative to the contribution of this annulus 
to the emission in all remaining annuli and vice versa. All annuli were therefore fitted simultaneously to remove the 
effect of the non-negligible PSF of EPIC, an effect that is more significant in the core regions, especially if there are strong 
gradients in the emission and spectral parameters over the field (as is the case of cool core clusters).

The spectra were analyzed with XSPEC v12.7.0 \citep{arnaud96} and fitted
with a single-temperature {\tt mekal} model \citep{liedahl95} with
Galactic absorption ({\tt wabs} model).
The fits were performed over the energy range $0.7-8.0$~keV for both 
of the two {\tt MOS} detectors, which were fitted simultaneously
to increase the statistics.
In our spectral fits, we left the temperature, the abundance,
and normalization free to vary. Local absorption was fixed to the Galactic neutral
hydrogen column density, as obtained from radio data  \citep{kalberla05}, 
and the redshift to the value measured from optical spectroscopy ($z$ in 
Table~\ref{exposures}). We used Cash statistics applied to the source plus
background\footnote{http://heasarc.gsfc.nasa.gov/docs/xanadu/xspec/manual/
XSappendixStatistics.html}, 
which requires a minimum grouping of the spectra (at least one count per spectral bin).
A goodness-of-fit for the best-fit models was computed with the 
XSPEC command {\tt goodness}, as described in BAL12.
In this way, we simulated for each spectrum in the sample 10,000 spectra from the best fit model, 
whose C-statistic values were compared with the value obtained from the original spectrum.
For each cluster, we found that the percentage of simulations with
a value of the C-statistic lower than that for the data is always below 20\%, 
an indication that the one-temperature model used to determine the temperature profiles
is always a good fit.

\subsection{Determination of $r_{500}$}\label{compr500}

Our analysis strategy, which involves testing for self-similarity in the temperature profiles at $z>0.4$, requires determining
the overdensity radius $r_{500}$ and the average temperature $T_{500}$ relative to that radius, computed as
described in BAL12, adopting the formula derived by \citet{vikhlinin06b}:
\begin{equation}
r_{500}hE(z)=0.792\left(\frac{T_{500}}{5\: keV}\right)^{0.53}\: Mpc,
\end{equation}
where $E(z)=(\Omega_m(1+z)^3+\Omega_\Lambda)^{0.5}$.

For each cluster, $T_{500}$ was estimated by extracting a spectrum
in the $0.15r_{500}<r<0.6r_{500}$ radial range, assuming a first-guess value of $T_{500}=5$~keV
to compute an initial value of $r_{500}$. The values of $T_{500}$ and $r_{500}$ were then
evaluated iteratively until a convergence to a stable value of the
temperature was obtained ($\Delta T_{500}\leq0.01$ keV between two successive
iterations).
The values of $T_{500}$ and $r_{500}$ for
each of the clusters are reported in Table~\ref{xrayprop}, which shows that
we sampled mostly high-temperature clusters;
$75\%$ of the clusters in the sample have a temperature $kT>6$~keV.


\section{Results}
\label{secresults}
\begin{figure}
\includegraphics[width=9.2cm, angle=0]{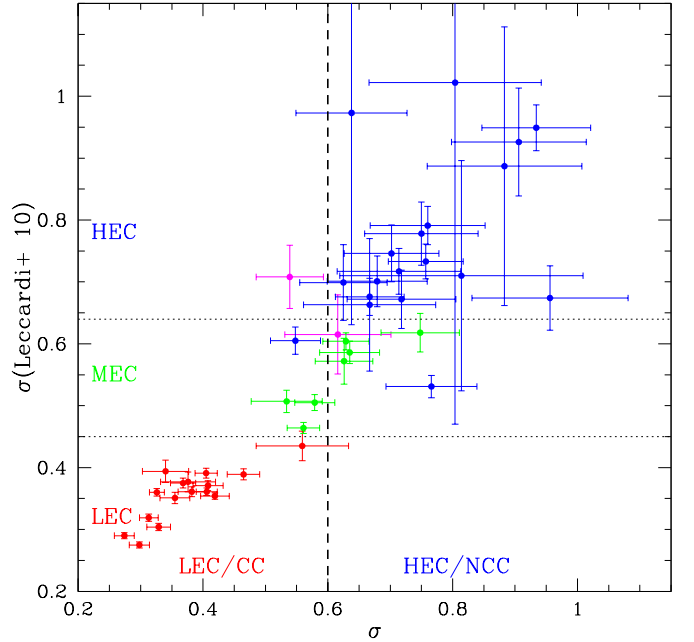}
\caption{Distribution of the values of the pseudo-entropy ratio $\sigma$ computed using the radial regions adopted in this paper vs. the values
of $\sigma$ computed using the radial range used in \citet{leccardi10} for the intermediate redshift cluster sample of \citet{leccardi08a}. 
The red, green, and blue squares represent the systems classified in the literature as CC, uncertain, and NCC , respectively. The purple squares
are the clusters classified in the literature as `dubious' CC.
The dotted horizontal lines represents the boundary chosen by \citet{leccardi10} to separate among LEC, MEC and HEC clusters.
The dashed thick vertical line represent the boundary we have chosen to distinguish between LEC/CC and HEC/NCC clusters. 
Evidently our classification
method for CC and NCC agrees very well with the literature and with the method of \citet{leccardi10}.}
\label{sigmacfr}
\end{figure}

\begin{figure*}
\includegraphics[width=18.0 cm, angle=0]{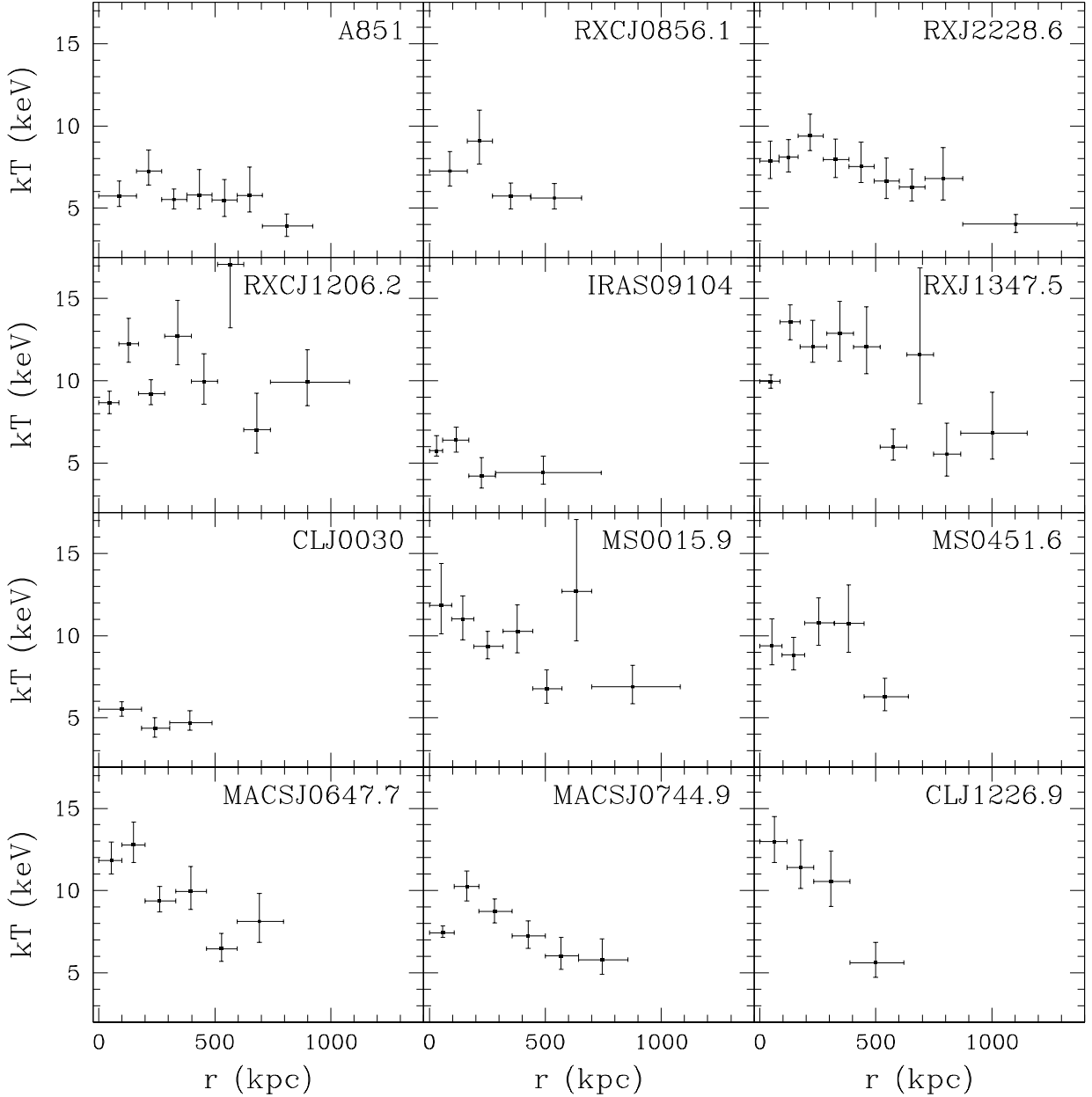}
\caption{Temperature profiles for the twelve galaxy clusters in our sample ordered according to redshift from left to
right and from top to bottom. The notes on individual clusters are reported in Appendix~\ref{individual}.}
\label{panels}
\end{figure*}

\subsection{CC/NCC classification of the clusters}\label{classif}

Although the spatial resolution of XMM and the redshift of the clusters in our sample
could in principle represent a complication and may not allow a trivial method to classify galaxy clusters, we divided our 
sample into CC and NCC clusters based on a comparison of the entropy measured in their core 
with the entropy measured outside the central regions. In particular, we considered the regions corresponding to 
$r<0.15r_{500}$ and to $0.15r_{500}\le r\le 0.4r_{500}$ , defining a pseudo-entropy ratio 
\begin{equation}
\sigma=(T_{IN}/T_{OUT})\times(EM_{IN}/EM_{OUT})^{-1/3}, 
\end{equation}
where $T_{IN}$ and $T_{OUT}$ are the temperature measured in the $r<0.15r_{500}$ region and in the $0.15r_{500}\le r\le 0.4r_{500}$ annulus,
respectively, and $EM_{IN}$ and $EM_{OUT}$ are the corresponding emission measures.
This ratio is very similar to the parameter defined in \citet{leccardi10} to classify their low-$z$ cluster sample ($z<0.24$), 
with the exception of the radial
range, which in their case was $r<0.05r_{200}$ for the IN region (corresponding to $\sim0.08r_{500}$) and $0.05r_{200}\le r\le 0.2r_{200}$
for the OUT region (corresponding approximately to $0.08r_{500}\le r\le 0.3r_{500}$). The choice of a different radial range was motivated
by the higher redhift of the clusters in our sample, which would make the IN region smaller ($\sim10^{\prime\prime}$) 
than the PSF of XMM-Newton, in most cases. To test whether this choice may introduce a bias in the classification, we computed the 
pseudo-entropy ratio for the cluster sample presented in \citet{leccardi08a} using both methods, and compared the results in Figure~\ref{sigmacfr}.
Clearly, the pseudo-entropy ratios calculated with the two different choices of radial range show a good
correlation with low dispersion especially at low values of $\sigma$.
In both methods, $\sigma$ is expected to be higher for NCC clusters and lower for CC clusters. \citet{leccardi10} divided
their sample into three different classes: low-entropy core (LEC, corresponding to $\sigma<0.45$), medium-entropy core (MEC, $0.45<\sigma<0.64$),
and high-entropy core (HEC, $\sigma>0.64$) clusters.
For our purposes, we do not need such a detailed classification, hence we decided to classify our clusters into only two categories:
the LEC/CC, with $\sigma<0.6$, and the HEC/NCC, with $\sigma\ge0.6$. 
Figure~\ref{sigmacfr} shows that the choice of $\sigma=0.6$ as a boundary between CC and NCC clusters is reasonable and agrees well with the 
independent methods of cluster classification in CC and NCC presented in \citet{ettori10}.
Our sample would then be divided into five CC clusters and seven NCC clusters, indicating a prevalence of non-relaxed NCC systems that is fully 
expected in our redshift range \citep[see e.g.][]{vikhlinin07,santos10}. The values of $\sigma$ for each cluster
are also reported in Table~\ref{xrayprop}.
In Figure~\ref{panels} we show the temperature profiles derived for each cluster in the sample with the analysis procedure
described above.

\subsection{Self-similarity of temperature profiles}\label{selfsim}

\begin{figure}
\includegraphics[width=9.2cm, angle=0]{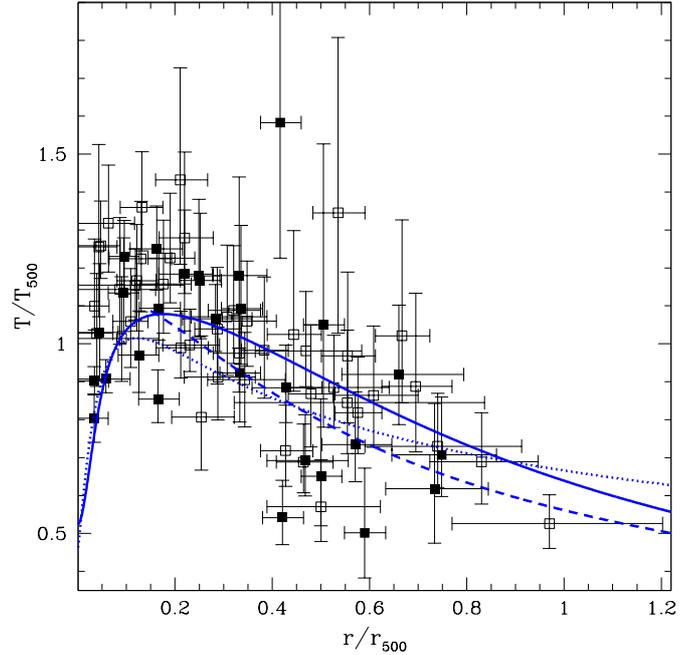}
\caption{Normalized temperature profiles for all clusters in the sample. CC clusters are plotted with filled squares, while
NCC clusters are shown as open squares (see \S~\ref{dichotomy}). The best fit to the \citet{vikhlinin06b} relation, freeing the
normalization, is drawn as a blue solid line. The blue dotted line represents the best fit to the relation
in Equation~\ref{fitvik06} (with $\xi_0=0.45$), while the blue dashed line is the best fit of Equation~\ref{powlaw} for $r>0.15r_{500}$.}
\label{kTnorm}
\end{figure}

\begin{figure}
\includegraphics[width=9.2cm, angle=0]{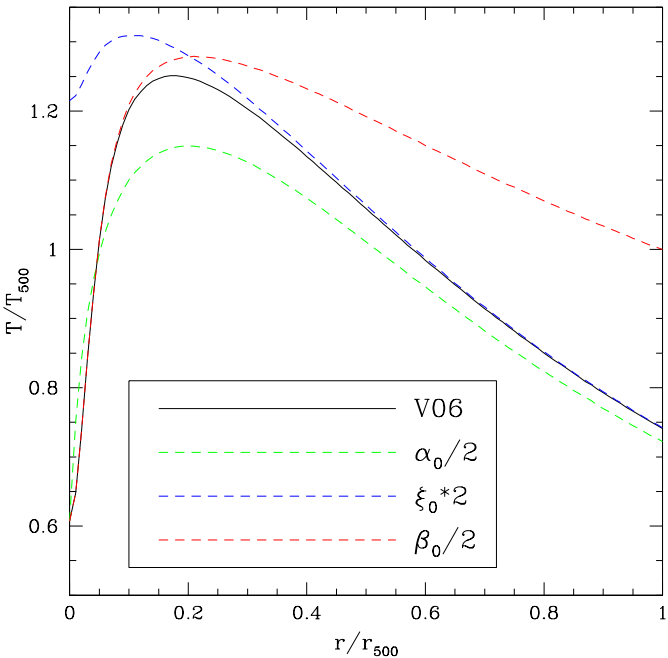}
\caption{Temperature profiles as described by the relation of \citet{vikhlinin06b} (solid black line) and by varying some of its parameters.
Decreasing the inner slope parameter $\alpha_0$ (green dashed line) creates a shallower temperature dip in the center. 
If the relative normalization $\xi_0$ between the inner and the outer profile is increased (blue dashed line), the profile transitions
toward the typical shape of NCC clusters. A decrease in the outer slope $\beta_0$ (red dashed line) flattens the external part
of the temperature profile.}
\label{V06behav}
\end{figure}

The values of $r_{500}$ and of $T_{500}$, computed in \S~\ref{compr500} and reported in Table~\ref{xrayprop},
were used to normalize the average distance from the center of each annulus and 
the corresponding observed temperature. The normalized temperature profiles are plotted in 
Figure~\ref{kTnorm}. At first glance, a self-similarity of the temperature profiles of the clusters in our sample can be
observed, although with non-negligible scatter. 
With the aim of quantifying the self-similarity, we considered a function derived from the relation 
introduced by \citet{vikhlinin06b} to fit the deprojected temperature profiles of their low-redshift CC cluster sample:
\begin{equation}
\label{fitvik06}
\frac{T(r)}{T_{500}}=A\frac{(x/a)^{\alpha_0}+\xi_0}{(x/a)^{\alpha_0}+1}\frac{1}{(1+(x/b)^2)^{\beta_0}},
\end{equation}
where $x=r/r_{500}$ and $A$ is a normalization parameter. 
The parameters $a$ and $b$ describe the core radius of the inner and outer regions of the profiles, respectively, while
$\alpha_0$ and $\beta_0$ represent the inner and the outer slope of the temperature profiles, respectively.
The additional parameter $\xi_0$ is introduced to adjust the relative normalization of the inner profile with respect to the outer profile.
Figure~\ref{V06behav} shows how each of the fitting parameters affects the shape of the temperature profiles described in 
Equation~\ref{fitvik06}.
The best-fit values of these parameters found by \citet{vikhlinin06b} were $A=1.35$, $a=0.045$, $b=0.6$, $\alpha_0=1.9$, $\beta_0=0.45$, 
and $\xi_0=0.45$. 
Because of the spatial resolution of XMM-Newton, the parameter $a$ is clearly insensitive to our data and it was therefore
fixed to the value of 0.045, defined by \citet{vikhlinin06b}. The other parameters are expected to be different by fitting 
our data since we are considering projected temperature profiles (and not 3-D profiles) of a sample not exclusively built of CC clusters. 
However, the best-fit values of these parameters obtained by \citet{vikhlinin06b} were used when we needed to freeze one parameter
to obtain meaningful constraints on the others.

For completeness, we also computed the projected equivalent of the \citet{vikhlinin06b} relation. The best-fit parameters of the projected
relation are slightly different from those derived for the 3-D profiles, with the most significant variation observed in the inner core radius $a$
(from 0.045 to 0.03), to which our data are not sensitive. We ran the fits of the temperature profiles using also the projected parameters when 
freezing one of the fit parameters became necessary, finding consistent results with respect to the fits performed using the corresponding 
3-D parameters. These findings show that a slight change in the value of a frozen parameter does not significantly affect the best fits of the 
others, therefore we can conclude that the results of the fits presented in the forthcoming sections are quite robust.

\begin{table*}
\caption{Values of $T_{500}$ and $r_{500}$ (expressed in Mpc and in arcsec, as computed in BAL12) for 
the {\em XMM-Newton} galaxy cluster bright sample presented in this paper. The effective radius $\bar{R}_{out}$ and the 
outer radius $R_{out}$ of the outermost radial bin are reported as well. The values of the pseudo-entropy
ratio $\sigma$ and the classification of each cluster are also shown. \label{xrayprop}}
\centering
\begin{tabular}{lrrrrrcl}
\hline
Cluster           & $T_{500}$ (keV)         & $r_{500}$ (Mpc) & $r_{500}$ ($^{\prime\prime}$) & $\bar{R}_{out}$ ($r_{500}$) & $R_{out}$ ($r_{500}$) & $\sigma$                   & Classification \\
\hline\hline
A851              &  $5.66\pm0.27$          & 0.975           & 180                           & 0.83                      & 0.95                & $0.896_{-0.179}^{+0.227}$  & NCC/HEC        \\
RXCJ0856.1+3756   &  $6.23\pm0.57$          & 1.023           & 187                           & 0.53                      & 0.64                & $0.723_{-0.172}^{+0.235}$  & NCC/HEC        \\
RXJ2228.6+2037    &  $7.60\pm0.42$          & 1.137           & 208                           & 0.97                      & 1.20                & $0.633_{-0.092}^{+0.100}$  & NCC/HEC        \\
RXCJ1206.2-0848   & $11.01_{-0.65}^{+0.60}$ & 1.361           & 239                           & 0.66                      & 0.79                & $0.413_{-0.055}^{+0.056}$  & CC/LEC         \\
IRAS09104+4109    &  $4.91_{-0.42}^{+0.43}$ & 0.886           & 155                           & 0.55                      & 0.84                & $0.582_{-0.126}^{+0.143}$  & CC/LEC         \\
RXJ1347.5-1145    & $11.23\pm0.43$          & 1.366           & 237                           & 0.73                      & 0.84                & $0.349_{-0.032}^{+0.037}$  & CC/LEC         \\
CLJ0030+2618      &  $4.69_{-0.30}^{+0.36}$ & 0.836           & 137                           & 0.47                      & 0.58                & $0.766_{-0.170}^{+0.207}$  & NCC/HEC        \\
MS0015.9+1609     &  $9.48_{-0.42}^{+0.57}$ & 1.184           & 186                           & 0.74                      & 0.91                & $0.877_{-0.131}^{+0.173}$  & NCC/HEC        \\
MS0451.6-0305     &  $9.10_{-0.53}^{+0.56}$ & 1.153           & 180                           & 0.47                      & 0.56                & $0.555_{-0.087}^{+0.101}$  & CC/LEC         \\
MACSJ0647.7+7015  &  $9.30_{-0.37}^{+0.45}$ & 1.138           & 171                           & 0.61                      & 0.70                & $0.700_{-0.079}^{+0.091}$  & NCC/HEC        \\
MACSJ0744.9+3927  &  $8.14_{-0.34}^{+0.34}$ & 0.995           & 139                           & 0.75                      & 0.86                & $0.451_{-0.048}^{+0.053}$  & CC/LEC         \\
CLJ1226.9+3332    & $10.16_{-0.73}^{+0.77}$ & 0.998           & 129                           & 0.50                      & 0.62                & $0.687_{-0.143}^{+0.183}$  & NCC/HEC        \\
\hline
\end{tabular}
\end{table*}

\subsubsection{Fits of CC and NCC temperature profiles}
\label{dichotomy}

The temperature profiles of galaxy clusters are known to be generally self-similar, showing a decline beyond $\sim15$\% of their
virial radius \citep[e.g.][]{vikhlinin05,baldi07,pratt07,leccardi08a}. However, an apparent difference is present in their center, 
with the relaxed CC clusters showing a temperature drop, while such drop is not observed in the NCC clusters, which are generally not 
relaxed.
Moreover, in a few studies based on low- and intermediate-redshift clusters, the NCC clusters were found to present a steeper temperature 
decline in the outer regions with respect to the CC clusters \citep[e.g.][]{baldi07}, although this still remains a debated question 
\citep[e.g.][]{leccardi08a}.

\begin{figure*}
\includegraphics[width=9.2cm, angle=0]{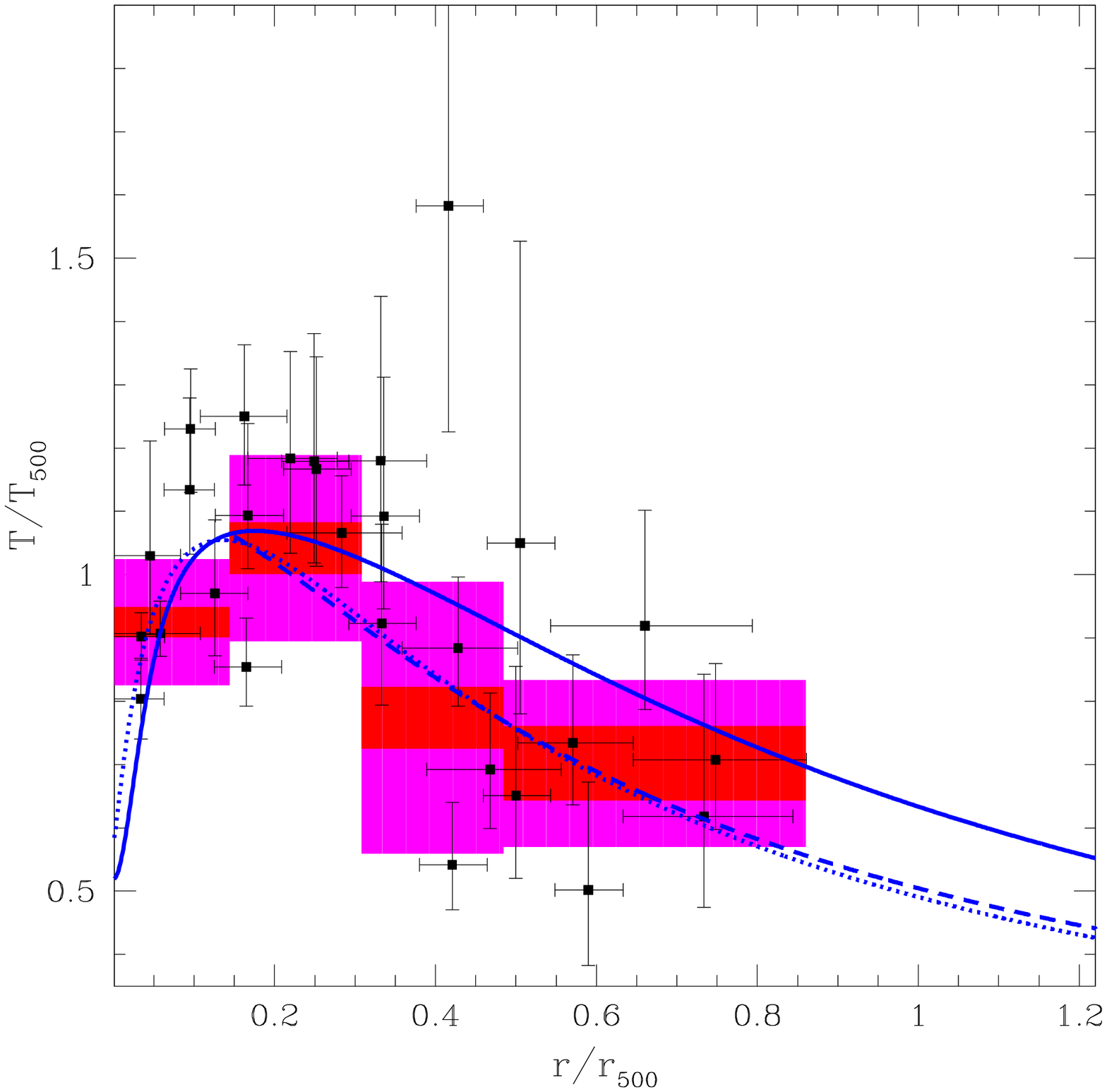}
\includegraphics[width=9.2cm, angle=0]{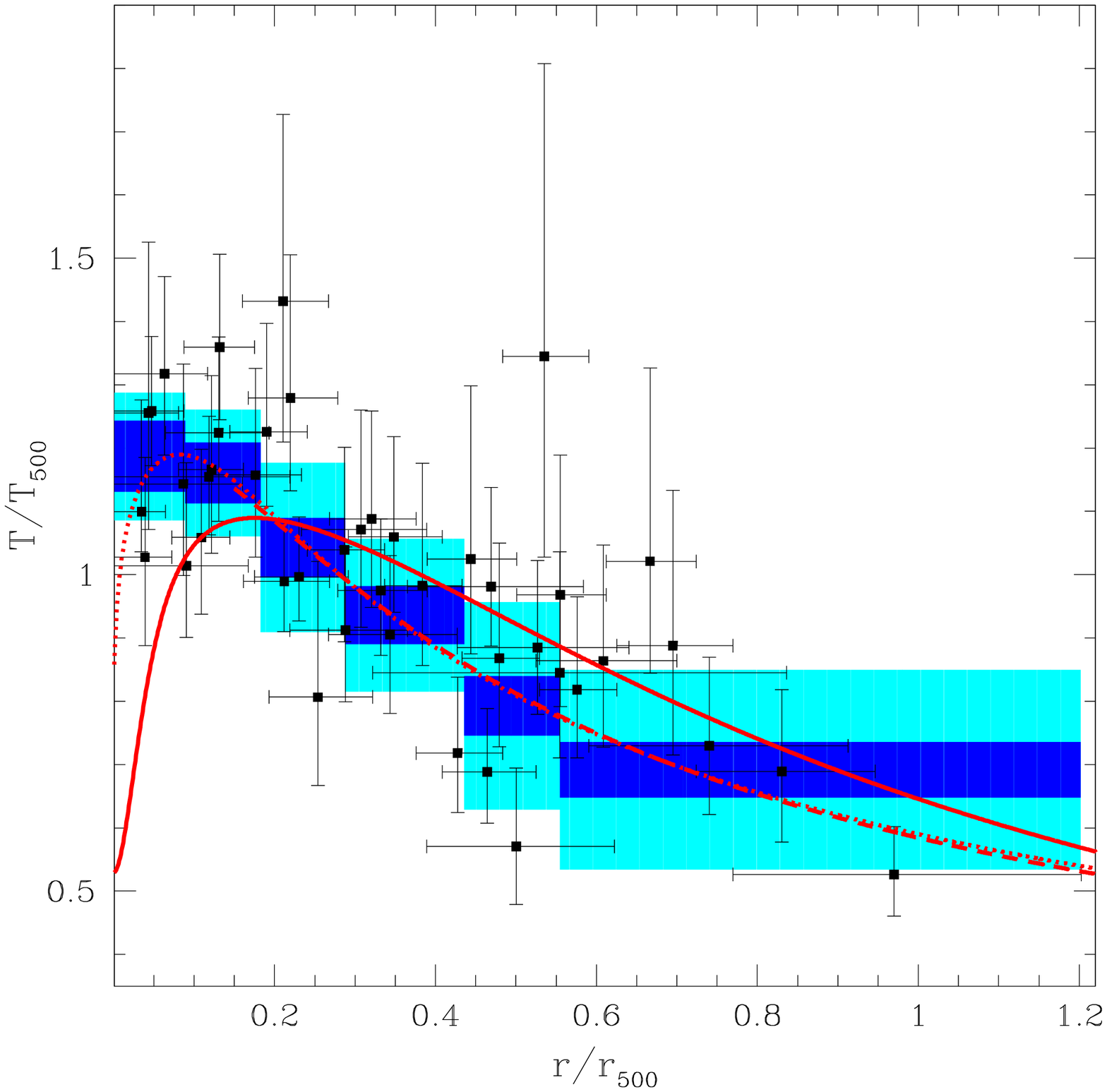}
\caption{{\it Left:} Normalized temperature profiles for the CC clusters in the sample. Shaded areas show the error-weighted mean of 
the normalized temperature with its error (red) and rms dispersion (magenta) in 4 radial bins. The best fit to \citet{vikhlinin06b}
relation freeing the normalization is plotted as a blue solid line, while the blue dotted line and the blue dashed
line represent the best fits to the relation in Equation~\ref{fitvik06} (with $\xi_0=0.45$) and to Equation~\ref{powlaw} (for $r>0.15r_{500}$), respectively.
{\it Right:} Normalized temperature profiles for the NCC clusters in the sample. Shaded areas show the error-weighted mean of 
the normalized temperature with its error (blue) and rms dispersion (cyan) in 6 radial bins. The best fit to \citet{vikhlinin06b} relation
(leaving the normalization free to vary), to the adaptation of the relation in Equation~\ref{fitvik06} (with $\xi_0=0.45$), and to Equation~\ref{powlaw} 
are plotted as a red solid, dotted and dashed line, respectively.}
\label{CCvsNCC}
\end{figure*}

As shown in \S~\ref{classif}, we divided our sample into CC and NCC based on their pseudo-entropy ratio $\sigma$. In
particular, we defined as CC clusters the objects having $\sigma<0.6$, while the objects with $\sigma\ge0.6$ were classified as 
NCC clusters. The two different categories were analyzed individually in this section. Figure~\ref{CCvsNCC} shows the normalized 
temperature profiles for CC and NCC clusters
separately. We also computed an emission-weighted average for both subsamples, which, as expected, shows the temperature drop typical 
of CC clusters in the central regions, while a flat (or increasing) temperature profile is observed at the center of NCC clusters.

To quantify the differences between the two subsamples, we fitted the relation in Equation~\ref{fitvik06} separately to CC and NCC 
cluster normalized temperature profiles (Table~\ref{CCvsNCCfit}). 
Running a fit leaving all the parameters in Equation~\ref{fitvik06} free to vary (except for $a$, which is insensitive to our data, 
and was therefore fixed to 0.045), no meaningful constraint could be obtained either on the CC or on the NCC cluster temperature profiles.
Hence, it was necessary to alternately freeze one of the parameters (except for the normalization $A$) to the value found in 
\citet{vikhlinin06b} (reported in \S~\ref{selfsim}), and allow the other parameters to fit the data.
For CC clusters, the better
fits are obtained by freezing the parameter $\xi_0$ or the parameter $\beta_0$, but in the latter case the value of the free parameter $\xi_0$ 
was frozen to zero by the fit. For the NCC clusters the better fits to Equation~\ref{fitvik06} are obtained when we freeze $\alpha_0$ or $\xi_0$.
If we compare the fits with the parameter $\xi_0$ frozen, which are showing one of the lowest $\chi^2$ and reasonable constraints
on the free parameters in both samples, CC and NCC clusters differ 
mainly for the best fit of the inner slope parameter $\alpha_0$, which assumes a positive value for the CC sample ($\alpha_0=1.14\pm0.59$), to fit the 
temperature drop in the center, and a value consistent with zero for the NCC sample ($\alpha_0=0.18\pm0.66$), to fit their flat temperature
profile in the central regions. 
No significant difference in the external slope $\beta_0$ could be detected, being $\beta_0=0.35\pm0.32$ in CC clusters and $\beta_0=0.30\pm0.10$ in 
NCC clusters.
Interestingly, the parameter $\xi_0$ in NCC clusters is always found to be $\sim1$ (when not frozen to 0.45). A value of $\xi_0\sim1$
makes the first term in Equation~\ref{fitvik06} approximately equal to unity, with the effect of removing the temperature dip in the center, as expected
in the profiles of NCC clusters. In the successive fits of NCC clusters presented in this paper we adopted a value of $\xi_0=1$.


\begin{table*}
\caption{Fit results using the functions defined in \S~\ref{selfsim}, obtained separately for the CC and NCC cluster sample, 
by fixing $a=0.045$ and by alternately freezing each one of the other free parameters, except the normalization $A$. 
The parameter $\alpha_0$ defines the inner slope, $\xi_0$ defines the relative normalization between the inner and outer regions of the profiles,
$b$ defines the core radius of the external regions, while $\beta_0$ defines the outer slope.
\label{CCvsNCCfit}}
\centering
\begin{tabular}{|c|l|c|ccccc|}
\hline
                     & Fit function                        & $\chi^2$/d.o.f. & $A$           & $\alpha_0$           & $\xi_0$              & $b$           & $\beta_0$           \\
\hline
\multirow{5}{*}{CC}  & Eq.~\ref{fitvik06} ($\alpha_0=1.9$) & 56.1/28         & $1.17\pm0.12$ & -                    & $0.61\pm0.15$        & $0.33\pm0.37$ & $0.35\pm0.40$       \\
                     & Eq.~\ref{fitvik06} ($\xi_0=0.45$)   & 55.8/28         & $1.30\pm0.05$ & $1.14\pm0.59$        & -                    & $0.28\pm0.26$ & $0.35\pm0.32$       \\
                     & Eq.~\ref{fitvik06} ($b=0.6$)        & 57.0/28         & $1.13\pm0.17$ & $1.89\pm2.76$        & $0.65\pm0.28$        & -             & $0.72\pm0.29$       \\
                     & Eq.~\ref{fitvik06} ($\beta_0=0.45$) & 55.8/29         & $1.88\pm0.07$ & $0.41\pm0.16$        & $0$\tablefootmark{a} & $0.30\pm0.07$ & -                   \\
\hline
\multirow{5}{*}{NCC} & Eq.~\ref{fitvik06} ($\alpha_0=1.9$) & 35.2/36         & $1.24\pm0.12$ & -                    & $0.94\pm0.24$        & $0.28\pm0.17$ & $0.29\pm0.12$       \\
                     & Eq.~\ref{fitvik06} ($\xi_0=0.45$)   & 35.2/36         & $1.65\pm0.09$ & $0.18\pm0.66$        & -                    & $0.26\pm0.19$ & $0.30\pm0.10$       \\
                     & Eq.~\ref{fitvik06} ($b=0.6$)        & 37.0/36         & $1.17\pm0.04$ & $0$\tablefootmark{a} & $0.94\pm0.13$        & -             & $0.62\pm0.08$       \\
                     & Eq.~\ref{fitvik06} ($\beta_0=0.45$) & 35.4/36         & $1.20\pm0.04$ & $unconstr.$          & $0.91\pm0.12$        & $0.44\pm0.05$ & -                   \\
\hline
\end{tabular}
\tablefoot{
\tablefoottext{a}{Parameter frozen by the fit}
}\end{table*}

\begin{table*}
\caption{Fit results using the functions similar to \citet{vikhlinin06b}, defined in \S~\ref{selfsim}, obtained by fixing $a=0.045$ and
by alternately freezing each one of the other free parameters, except the normalization $A$, which was always left free to vary.
The meaning
of the fit parameters is the same as in Table~\ref{CCvsNCCfit}. The parameters $\eta$ and $\zeta$ take into
account the dependence of the profiles on the pseudo-entropy ratio $\sigma$ and on the redshift $z$, respectively. \label{fitV06}}
\centering
\begin{tabular}{|l|c|ccccc|cc|}
\hline
Fit function                                                & $\chi^2$/d.o.f. & $A$           & $\alpha_0$     & $\xi_0$        & $b$           & $\beta_0$     & $\eta$         & $\zeta$        \\
\hline
Eq.~\ref{fitvik06} ($\alpha_0=1.9$)                         & 107.4/68        & $1.25\pm0.09$ & -              & $0.56\pm0.11$  & $0.25\pm0.13$ & $0.28\pm0.11$ & -              & -              \\
Eq.~\ref{fitvik06} ($\xi_0=0.45$)                           & 107.2/68        & $1.35\pm0.04$ & $1.39\pm0.46$  & -              & $0.22\pm0.09$ & $0.28\pm0.10$ & -              & -              \\
Eq.~\ref{fitvik06} ($b=0.6$)                                & 111.1/68        & $1.13\pm0.05$ & $3.14\pm2.47$  & $0.72\pm0.10$  & -             & $0.61\pm0.10$ & -              & -              \\
Eq.~\ref{fitvik06} ($\beta_0=0.45$)                         & 109.2/68        & $1.17\pm0.10$ & $2.45\pm2.13$  & $0.66\pm0.15$  & $0.44\pm0.08$ & -             & -              & -              \\
\hline
Eq.~\ref{fitvik06sigmaz} ($b=0.4$; $\xi_0=0.45$; &\multirow{2}{*}{109.7/68}   & \multirow{2}{*}{$1.17\pm0.06$} & \multirow{2}{*}{$4.78\pm5.62$}  & \multirow{2}{*}{-}              & \multirow{2}{*}{-}             & \multirow{2}{*}{$0.26\pm0.09$} & \multirow{2}{*}{$0.33\pm0.18$}  & \multirow{2}{*}{-}             \\
 $\:\:\:\:\:\:\:\:\:\:\:\:\zeta=0$) & & & & & & & & \\
Eq.~\ref{fitvik06sigmaz} ($b=0.4$; $\xi_0=0.45$)            & 110.3/67        & $1.48\pm0.15$ & $0.61\pm0.44$  & -              & -             & $0.64\pm0.49$ & $-0.22\pm0.20$ & $0.12\pm0.39$  \\
\hline
\end{tabular}
\end{table*}

\subsubsection{General fits of the whole cluster sample}\label{genfits}

We fit the relation in Equation~\ref{fitvik06} to the normalized temperature profiles of the whole sample, including both CC and NCC clusters. 
Leaving all fit parameters free to vary (except for $a$, which is insensitive to our data, and was therefore fixed to 0.045), we obtain that 
three of them ($A$, $\alpha_0$ and $b$) are unconstrained and able to vary throughout the whole parameter space without introducing any
difference in the value of the statistics,
with the other two being loosely constrained ($\xi_0=0.19\pm0.21$, $\beta_0=0.35\pm1.08$). 
Therefore, as we did for the separate fits of CC and NCC clusters (\S~\ref{dichotomy}), we alternately froze one of the parameters 
(except the normalization $A$) 
to the value found by \citet{vikhlinin06b} (reported in \S~\ref{selfsim}), and allowed the other four parameters to fit the data.
The results of the fits are presented in Table~\ref{fitV06}.
The better fits were obtained when we froze $\alpha_0$ or $\xi_0$ (with a minimum difference in their respective $\chi^2$),
yielding a best-fit value of $\xi_0=0.56\pm0.11$ and $\alpha_0=1.39\pm0.46$, respectively. These fits also give
fully consistent values of the normalization $A$ and of the parameters $b$ and $\beta_0$, which describe the external region of the 
profiles.

\begin{figure}
\includegraphics[width=9.2cm, angle=0]{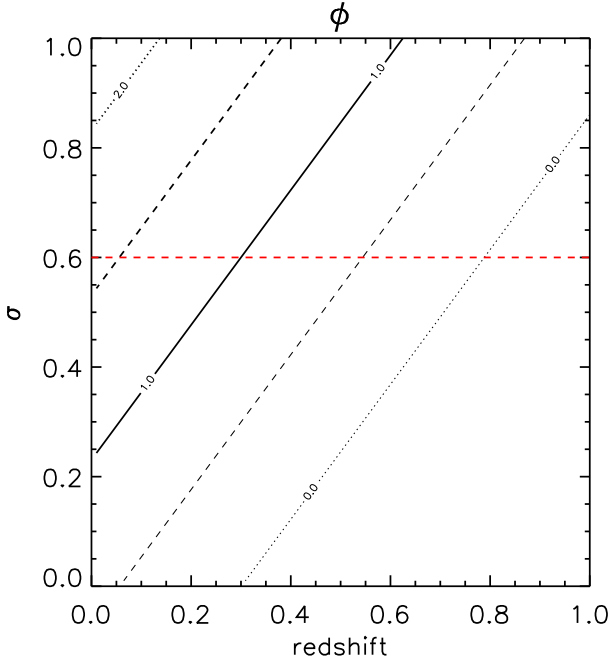}
\caption{Expected values for the function $\phi_+$ introduced in Equation~\ref{fitvik06sigmaz} in the $z-\sigma$ parameter space. 
The black lines represent the loci where a combination of the values of $z$ and $\sigma$ are expected to yield 
identical values of $\phi_+$. The dashed red line represent the boundary between CC and NCC clusters.}
\label{phi}
\end{figure}

We also adopted a more general approach to test the dependence of the profiles on the pseudo-entropy ratio $\sigma$ 
(that we used to classify the clusters in our sample as CC or NCC), and/or on the cluster redshift, by fitting the data with a function of the distance 
from the center $r$, the pseudo-entropy ratio $\sigma$, and the redshift $z$. Equation~\ref{fitvik06} was modified accordingly in the form
\begin{equation}
\label{fitvik06sigmaz}
\left\{\begin{array}{l}
\displaystyle \frac{T}{T_{500}}=A\frac{(x/0.045)^{\alpha}+\xi}{(x/0.045)^{\alpha}+1}\cdot\frac{1}{(1+(x/b)^2)^{\beta}},\\
\\
\displaystyle   \xi = \xi_0 \cdot \phi_+(\sigma,\zeta=0),\\
\\
\displaystyle   \alpha = \alpha_0 \cdot \phi_-(\sigma,z),\\
\\
\displaystyle   \beta = \beta_0 \cdot \phi_+(\sigma,z),\\
\\
\displaystyle   \phi_\pm(\sigma,z) = 1\pm\eta\cdot(1+\sigma)\pm\zeta\cdot(1+z)\:. \end{array} \right.
\end{equation} 
In the function $\phi_\pm(\sigma,z)$, the parameters $\eta$ and $\zeta$
take into account the dependence of the temperature profiles on the pseudo-entropy ratio $\sigma$ and 
on the redshift $z$, respectively (Figure~\ref{phi}). Owing to the expected evolution of the temperature profiles in CC clusters \citep[e.g.][]{ettori08},
$\beta$ is expected to increase at high redshift in CC systems and therefore we considered the function $\phi_+$ to model
their evolution. On the other hand, $\alpha$ is expected to decrease with redshift and the function $\phi_-$ was considered for the
evolution of this parameter.
Although in principle $\xi$ is expected to have a dependence on $z$, no meaningful constraint could be obtained for its evolution and therefore 
we imposed this parameter to have a dependence only on $\sigma$ (by setting $\zeta=0$ in the function $\phi_+$).

Initially, we tested the dependence of the temperature profiles only on the parameter $\sigma$ (an indicator of their state of relaxation) 
by fitting Equation~\ref{fitvik06sigmaz} to the profiles and considering no redshift dependence at this stage (i.e. fixing the parameter $\zeta=0$). 
Similarly to the fits of Equation~\ref{fitvik06}, we froze one of the free parameters to obtain reasonable constraints on the others.
In particular, we chose to freeze the value of $\xi_0=0.45$, corresponding to the case with the lowest value of $\chi^2$. This fit, however, 
is not able
to meaningfully constrain any of the free parameters. Therefore we decided to also fix the value of the external core radius to $b=0.4$, a 
value consistent with the results of the fits in Table~\ref{fitV06}.
The fit yields a $\chi^2\sim109.7$ with 68 d.o.f. and points out a mild dependence on $\sigma$ of the temperature profiles, as indicated by the best-fit 
value of $\eta=0.33\pm0.18$
(Table~\ref{fitV06}). 
Even taking into account the uncertainties in the measurement of $\eta$ and of the other parameters, the 
outer part of the profiles clearly becomes steeper for higher values of $\sigma$ (i.e. transitioning toward the NCC clusters).
This behavior agrees with the results obtained in lower redshift galaxy cluster samples \citep[e.g.][]{baldi07} and with
the universal pressure profile derived by \citet{arnaud10} in galaxy clusters in the REXCESS local sample ($z<0.2$), 
showing a low dispersion, especially in the external regions. 
An analogous behavior of the pressure profiles for both CC and NCC clusters, combined with that of the density profiles, which are known to be
flatter at large radii in NCC clusters than in CC clusters \citep[see e.g.][]{cavaliere09,eckert11}, would indeed yield steeper temperature 
profiles in NCC than in CC systems.

\begin{table*}
\caption{Fit results for the external regions of the temperature profiles ($r>0.15r_{500}$) using the power-law functions defined in
\S~\ref{external}. The parameters $A$ and $\beta_0$ describe the normalization and the slope of the power-law, respectively. Also in this case,
$\eta$ and $\zeta$ are introduced to take into account the dependence of the profiles on $\sigma$ and $z$, respectively.
\label{fitexternal}}
\centering
\begin{tabular}{|l|c|c|cc|cc|}
\hline
Fit function                                         & $\chi^2$/d.o.f. & F-test prob.  & $A$            & $\beta_0$     & $\eta$         & $\zeta$       \\
\hline
Equation~\ref{powlaw} ($b=0.4$, $\eta=0$, $\zeta=0$) & 73.2/51         & -             & $1.13\pm0.05$  & $0.37\pm0.06$ & -              & -             \\
Equation~\ref{powlaw} ($b=0.4$, $\zeta=0$)           & 68.3/50         & 96\%          & $1.14\pm0.05$  & $0.95\pm0.32$ & $-0.37\pm0.08$ & -             \\
Equation~\ref{powlaw} ($b=0.4$)                      & 67.6/49         & 91\%          & $1.14\pm0.05$  & $0.64\pm0.54$ & $-0.53\pm0.33$ & $0.30\pm0.64$ \\
\hline
\end{tabular}
\end{table*}

\begin{table*}
\caption{Fit results for the external regions of the temperature profiles ($r>0.15r_{500}$) using the functions defined in \S~\ref{external}, 
obtained separately for the CC and NCC cluster samples. The fit parameters have the same meaning as in Table~\ref{fitexternal}. 
Again, the dependence on $\sigma$ (expressed by $\eta$) is not considered here since it was already used to subdivide the sample between CC and NCC.
\label{CCvsNCCfitexternal}}
\centering
\begin{tabular}{|c|l|c|c|cc|c|}
\hline
                     &                                                      & $\chi^2$/d.o.f. & F-test prob.  & $A$           & $\beta_0$     & $\zeta$        \\
\hline 
\multirow{2}{*}{CC}  & Equation~\ref{powlaw} ($b=0.4$, $\eta=0$, $\zeta=0$) & 42.2/21         & -             & $1.11\pm0.08$ & $0.39\pm0.12$ & -              \\
                     & Equation~\ref{powlaw} ($b=0.4$, $\eta=0$)            & 38.5/20         & 82\%          & $1.12\pm0.08$ & $1.66\pm0.92$ & $-0.48\pm0.09$  \\
\hline
\multirow{2}{*}{NCC} & Equation~\ref{powlaw} ($b=0.4$, $\eta=0$, $\zeta=0$) & 28.2/28         & -             & $1.17\pm0.06$ & $0.37\pm0.06$ & -              \\
                     & Equation~\ref{powlaw} ($b=0.4$, $\eta=0$)            & 24.4/28         & -             & $1.19\pm0.05$ & $0.06\pm0.01$ & $4$\tablefootmark{a} \\
\hline
\end{tabular}
\tablefoot{
\tablefoottext{a}{Parameter frozen to the upper limit allowed by the fit}
}\end{table*}

We also added a redshift dependence of the temperature profiles by fitting Equation~\ref{fitvik06sigmaz} to the data, this time also leaving
the parameter $\zeta$ free to vary.
Fixing again $\xi_0=0.45$ and $b=0.4$, 
we obtain a $\chi^2=110.3$ with 67 degrees of freedom, i.e. no improvement in the fit quality. The best-fit value of $\zeta$ is found to be consistent with zero 
($\zeta=0.12\pm0.39$), pointing to no significant evolution with the redshift within the redshift range considered ($0.4<z<0.9$). In this fit, 
the parameter $\eta$ is also consistent with zero, an indication
that also introducing a redshift dependence removed the mild dependence of the profiles on $\sigma$ detected in the previous fit.

We caution, however, that our sample is not a complete sample at $z>0.4$ with representative fractions of CC and NCC clusters (\S~\ref{sampsel}) 
and the value of the parameter $\eta$ may change if we use a different cluster sample than this present one.

\subsubsection{Fits of the cluster external regions}\label{external}

As shown before, the shape of temperature profiles in galaxy clusters is not easy to describe by using a simple function with few
free parameters, to fit the observed data.
Indeed, in the previous subsections the relations we considered have at least five free parameters, which makes it harder to
obtain meaningful constraints on the best-fit values and complicates the interpretation of the resulting best-fit functions.
The modeling of the temperature profiles can be simplified if we consider only the external regions of the clusters 
($r>0.15r_{500}$) in a radial range in which the influence of a possible cool core is marginal.
To fit the profiles in these regions we considered Equation~\ref{fitvik06sigmaz} and imposed that the parameters $\alpha$ and $\xi$
are frozen to zero. Thuse we obtain
\begin{equation}
\label{powlaw}
\frac{T(r)}{T_{500}}=A\cdot\frac{1}{(1+(x/b)^2)^{\beta}},
\end{equation}
where the parameters $b$ and $\beta$ have the same meaning as in Equation~\ref{fitvik06sigmaz}.
To simplify the fit and have just two free parameters, we also fixed the value of the core radius $b$ to 
0.4, a value that is more appropriate to the best fits found previously in \S~\ref{genfits} for our sample 
than the value of 0.6 measured by \citet{vikhlinin06b} in their low-redshift CC clusters.
Without considering any dependence on $\sigma$ and/or $z$, the best fits to the data of Equation~\ref{powlaw} 
yield a $\chi^2=73.2$ (both with 51 d.o.f.) 
with best-fit values of the normalization $A=1.13\pm0.05$ and of $\beta_0=0.37\pm0.06$ (Table~\ref{fitexternal}). 

Introducing a dependence on $\sigma$ (i.e. leaving $\zeta=0$ in Equation~\ref{powlaw} and freeing the parameter $\eta$), a mild dependence 
on this parameter could be found, with the value of $\eta=-0.37\pm0.08$). Freeing also the parameter $\zeta$, no redshift dependence
could be found ($\zeta=0.30\pm0.64$) within our redshift range.

By separately fitting
the CC and NCC samples to Equation~\ref{powlaw} we obtained a $\chi^2/d.o.f.=42.2/21$ and a $\chi^2/d.o.f.=28.2/28$, respectively, indicating that 
the function is a good fit for NCC cluster profiles and pointing to more scattered profiles in CC clusters. 
Both profiles show consistent values of the external slope $\beta_0$ ($\beta_0=0.37\pm0.06$ for NCC, $\beta_0=0.39\pm0.12$ for CC clusters).

We also tested the dependence of the profiles of CC and NCC clusters on the redshift, considering the relation in Equation~\ref{powlaw} and fixing $\eta=0$ 
(Table~\ref{CCvsNCCfitexternal}). The fit yielded an improvement in the fit quality of CC clusters (F-test 
probability $\sim82$\%), while no F-test could be computed for the NCC clusters because the parameter $\zeta$ was frozen to the upper limit allowed
by the fit.
In CC clusters, we detect a significant dependence of the slope from the redshift, as shown by the value of $\zeta=-0.48\pm0.09$ significantly different from
zero. However, the error on the slope is quite large ($\beta_0=1.66\pm0.92$), preventing any quantitative conclusion on the evolution of the 
external temperature profiles
in CC clusters within the redshift range probed by our sample. 

\section{Discussion}

In \S~\ref{secresults}, we presented the normalized temperature profiles of our galaxy cluster sample, showing that they are generally self-similar at large radii
($r>0.15r_{500}$). The profiles exhibit a clear decline of the temperature with the radius, although a non-negligible scatter is present, especially among the CC 
clusters.
The function we used to fit the profiles \citep[derived from][]{vikhlinin06b} represents a good fit to the data in most cases, with the largest $\chi^2$ observed
for the CC sample (where the scatter is larger). Although the profiles showed some dependence from the pseudo-entropy ratio $\sigma$ (that we used to classify the clusters
based on their state of relaxation), dividing the sample into CC and NCC cluster, no difference in the external slope between CC and NCC clusters was found, in 
contrast with some of the previous works performed at lower redshift \citep[e.g.][]{degrandi02,baldi07}. Moreover, an attempt to introduce a dependence also on 
the redshift gave inconclusive results. These facts may be mostly due to the sample size, which consists of 12 galaxy clusters 
(five CC and seven NCC clusters), for a total 
of $\sim$70 data points ($\sim$30 and $\sim$45 data points for the CC and NCC sample, respectively). Moreover, a reduced redshift range ($0.4<z<0.9$) may factor in a
lack of evidence of temperature profile evolution within our sample.

Both of these problems could be tackled considering a larger sample of clusters, spanning a different redshift range. To this aim, the most
suitable sample 
clearly is the sample of 44 XMM-Newton clusters presented in \citet{leccardi08a}, which spans a lower redshift range than our sample ($0.1<z<0.3$) and whose data 
were reduced and analyzed using a procedure similar to that adopted in this paper, especially concerning the XMM background treatment.

\begin{figure}
\includegraphics[width=9.2cm, angle=0]{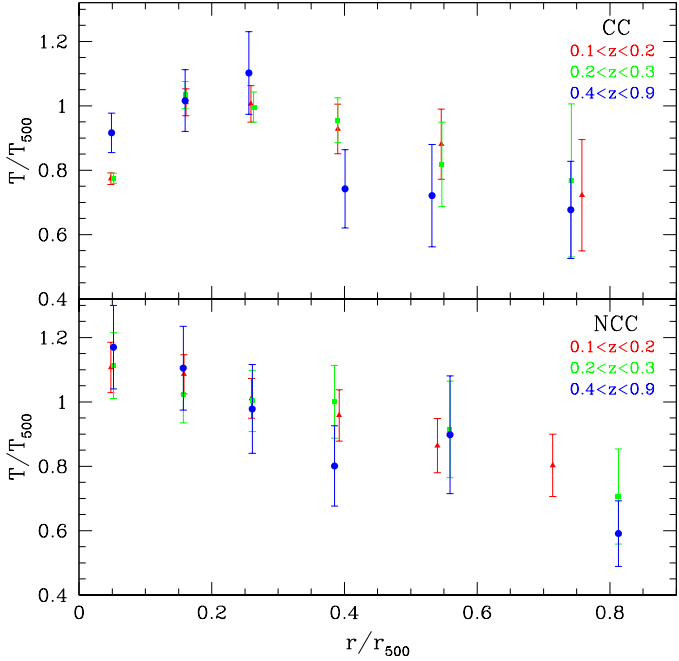}
\caption{{\it Top:} Average temperature profiles measured for the CC clusters presented in this paper and for the CC clusters of \citet{leccardi08a}, 
examined in three redshift bins: $0.1<z<0.2$ (red triangles), $0.2<z<0.3$ (green squares), and $0.4<z<0.9$ (blue dots). The profiles are 
consistent to be
similar except for the central radial bin where the temperature dip in the high-redshift clusters is less pronounced. High-redshift clusters are indeed
expected to be at a less advanced stage of their formation and are expected to have not grown enough mass to create a deep potential well 
and establish an efficient 
cooling in the central regions. We caution, however, that another concurring factor may be the difficulty to fully resolve the core at high redshift.
{\it Bottom:} Average temperature profiles measured for the NCC clusters presented in this paper and for the NCC clusters of \citet{leccardi08a}, 
examined in three redshift bins: $0.1<z<0.2$ (red triangles), $0.2<z<0.3$ (green squares), and $0.4<z<0.9$ (blue dots). Although the average profiles are consistent
at every radius, the slope of the high-redshift sample seems slightly steeper than that of the lower redshift samples, which is fully expected because of the
less advanced stage of formation (and consequently less massive hot gas halos) of higher redshift clusters.}
\label{avgprof_lec_bal}
\end{figure}

\subsection{Comparison with intermediate redshift clusters}\label{compint}

The normalized temperature profiles for the sample of \citet{leccardi08a} were obtained
using the values of $r_{500}$ reported in \citet{ettori10} and the values of $T_{500}$ were computed using emission-weighted averages of the temperatures 
measured in the radial bins comprised in the 0.15-0.6 $r_{500}$ radial range.
We computed the emission-weighted average temperature profiles of CC and NCC clusters in different radial bins to achieve a direct comparison 
between the two samples
(probing different redshift ranges). Since the size of the \citet{leccardi08a} sample is quite large (44 clusters), we divided it into two
subsamples, considering two redshift bins: $0.1<z<0.2$ and $0.2<z<0.3$. The $0.4<z<0.9$ redshift range is covered by the sample presented in this work.
The emission-weighted average temperature profiles in the three redshift bins are shown in Figure~\ref{avgprof_lec_bal}. 

The profiles of CC clusters are consistent within the errors
at every radius except for the central radial bin where the high-redshift cluster sample shows a shallower temperature dip with
respect to the lower redshift clusters. 
This may indicate a difference between CC clusters at $z>0.4$ with respect to CC clusters located at $z<0.3$. A possible reason for this is that
the former are at less advanced
stage of their formation and would not have had enough time to accrete enough mass to create a deeper potential well and therefore establish a more efficient 
cooling in the central regions.
However, in this case we have to be cautious about any claim about evolution for two reasons. First of all, it may be very
difficult to fully resolve the cluster core regions at $z>0.4$ because of the XMM-Newton PSF. Although a PSF correction was applied (\S~\ref{psf}),
the size of the region corresponding to the central bin ($r<0.1r_{500}$) in some cases could be about that of the XMM-Newton PSF itself. 
Moreover, selection effects could affect this result. Indeed, our high-redshift sample consists of only five clusters and cannot be considered complete 
in any sense (\S~\ref{sampsel}).

\begin{table*}
\caption{Fit results for the CC clusters in our sample and in the \citet{leccardi08a} sample considering three different redshift bins 
and using the relation in Equation~\ref{fitvik06} obtained by fixing the value of the relative normalization
between the internal and external region of the profiles to $\xi_0=0.45$ and the values of the external core radius to $b=0.4$. The meanings
of the fit parameters are the same as in Table~\ref{fitV06}. \label{comp_LEC_BAL_CC}}
\centering
\begin{tabular}{|l|c|ccc|}
\hline
Redshift range    & $\chi^2$/d.o.f. & $A$           & $\alpha_0$    & $\beta_0$     \\
\hline
$0.1<z<0.2$       & 561.7/94        & $1.14\pm0.01$ & $2.33\pm0.25$ & $0.31\pm0.05$ \\
$0.2<z<0.3$       & 186.2/51        & $1.02\pm0.02$ & $1.96\pm0.40$ & $0.09\pm0.05$ \\
$0.4<z<0.9$       & 56.2/29         & $1.29\pm0.04$ & $1.01\pm0.39$ & $0.51\pm0.09$ \\
\hline
\end{tabular}
\end{table*}
\begin{table*}
\caption{Fit results for the external regions of the temperature profiles ($r>0.15r_{500}$) using the function of Equation~\ref{powlaw} (defined in \S~\ref{external})
obtained separately for the CC and NCC cluster samples in three different redshift bins. The fit parameters have the same meaning as in Table~\ref{fitexternal}.
\label{comp_powlaw_BAL_LEC}}
\centering
\begin{tabular}{|c|c|c|cc|}
\hline
                     & Redshift range & $\chi^2$/d.o.f. & $A$           & $\beta_0$    \\
\hline
\multirow{3}{*}{CC}  & $0.1<z<0.2$    & 61.4/66         & $1.09\pm0.01$ & $0.24\pm0.03$ \\
                     & $0.2<z<0.3$    & 32.4/39         & $1.07\pm0.02$ & $0.18\pm0.03$ \\
                     & $0.4<z<0.9$    & 42.2/21         & $1.11\pm0.08$ & $0.39\pm0.12$ \\
\hline
\multirow{3}{*}{NCC} & $0.1<z<0.2$    & 52.4/57         & $1.09\pm0.02$ & $0.20\pm0.03$ \\
                     & $0.2<z<0.3$    & 50.5/52         & $1.08\pm0.03$ & $0.16\pm0.04$ \\
                     & $0.4<z<0.9$    & 28.2/28         & $1.17\pm0.06$ & $0.37\pm0.06$ \\
\hline
\end{tabular}
\end{table*}

For the NCC clusters, the average profiles in the three redshift bins are consistent within the errors at every radius. It is worth noticing, however, 
that the slope of our high-redshift sample seems to be slightly steeper than that of the lower redshift samples, although no quantitative claim can be obtained
from this plot. In this case the PSF of XMM is not a biasing factor, but selection effects may also play a role in NCC clusters (\S~\ref{sampsel}).

A more quantitative comparison could be performed by considering the relations introduced in \S~\ref{selfsim} to fit the data points from the samples in the
three different redshift bins. For the CC clusters, where the main differences are observed in the central regions, the most suitable comparison can
be made using Equation~\ref{fitvik06} and fixing the value of the relative normalization
between the internal and external region of the profiles to $\xi_0=0.45$. Since we are not interested in the external regions of the profiles, at this stage
we can also fix the external core radius to $b=0.4$. 
The results of the fits for the three different redshift bins are shown in Table~\ref{comp_LEC_BAL_CC}.
The inner slope $\alpha_0$ is flatter in the high-z sample presented in this paper with respect to the two lower redshift samples extracted
from \citet{leccardi08a}, although the difference is statistically significant only in the comparison with the $0.1<z<0.2$ redshift bin ($>2\sigma$).
It is worth noticing, however, that also this comparison could be biased by the XMM PSF
and by possible selection effects. Moreover, there is a significant difference in the external slope $\beta_0$, but no quantitative claim could be made about
this point because fixing the external core radius to $b=0.4$ may have biased the determination of the slope, especially at lower redshift.

In the NCC clusters the average profiles are showing a slightly steeper slope in our high-z sample with respect to the intermediate cluster sample of 
\citet{leccardi08a}.
The difference could be quantified by fitting the data points for both samples using a simplified function as described in \S~\ref{external} 
(Equation~\ref{powlaw}) 
and considering only the external regions of the clusters ($r>0.15r_{500}$). We used this function also to compare the external slopes at different 
redshifts in CC clusters, a task not attainable with the more complex function in Equation~\ref{fitvik06}.
The results of the power-law fits are shown in Table~\ref{comp_powlaw_BAL_LEC}.
\begin{table*}
\caption{Fit results using the functions similar to \citet{vikhlinin06b} (defined in \S~\ref{selfsim}) for our sample and the \citet{leccardi08a} sample
together, obtained by fixing $a=0.045$ and
the relative normalization $xi_0=0.45$. The meaning of the fit parameters is the same as in Table~\ref{fitV06}. \label{fitV06all}}
\centering
\begin{tabular}{|l|c|cccc|cc|}
\hline
Fit function                                       & $\chi^2$/d.o.f. & $A$           & $\alpha_0$    & $\xi_0$       & $\beta_0$      & $\eta$         & $\zeta$        \\
\hline
Eq.~\ref{fitvik06} ($b=0.4$)                       & 1719.0/384      & $1.15\pm0.01$ & $1.66\pm0.35$ & $0.41\pm0.06$ & $0.49\pm0.08$  & -              & -              \\
\hline
Eq.~\ref{fitvik06sigmaz} ($b=0.4$; $\zeta=0$)      & 1663.9/383      & $1.13\pm0.04$ & $3.13\pm1.04$ & $0.32\pm0.05$ & $0.17\pm0.04$  & $0.29\pm0.11$  & -              \\
Eq.~\ref{fitvik06sigmaz} ($b=0.4$)                 & 1662.6/382      & $1.14\pm0.05$ & $3.85\pm1.79$ & $0.33\pm0.06$ & $0.16\pm0.04$  & $0.23\pm0.11$  & $0.17\pm0.09$  \\
\hline
\end{tabular}
\end{table*}
For the CC clusters there is no significant difference in the power-law slope between the two intermediate redshift bins, where the value of $\beta$
is consistent within the 1$\sigma$ errors. A difference just above 1$\sigma$ is observed instead between the slope of the high-z CC clusters 
and the slope of the intermediate redshift CC clusters.
For the NCC no difference in the slope $\beta$ is observed either between the two intermediate redshift bins.
The value of $\beta$ in our high-z NCC sample is in this case significantly higher than in the NCC clusters at $z<0.3$ in the \citet{leccardi08a} 
sample ($\sim$2$\sigma$). As stated before, this difference could be caused by selection effects introduced by our sample selection. However,
it is very likely that it is instead related to the different stages of the evolution of the clusters at $z>0.4$ with respect to clusters at $z<0.3$. 
It is indeed expected that the former would be younger clusters with a smaller sized hot gas halo with respect to the latter, 
and therefore would show lower temperatures at large radii.

\subsection{A universal law for temperature profiles}

The data points of the clusters in the \citet{leccardi08a} sample can be used together with our sample to help us define a universal behavior of temperature
profiles in galaxy clusters at $0.1<z<0.9$ as a function of both cosmic time and state of relaxation (as defined by the pseudo-entropy ratio $\sigma$).
The relations introduced in \S~\ref{selfsim} could indeed be used to jointly fit the data points from our high-z cluster sample and the intermediate-redshift
sample of \citet{leccardi08a}.
The results are presented in Table~\ref{fitV06all}. For simplicity, we considered only the case of Equation~\ref{fitvik06} where the value of the 
external core radius has been set at $b=0.4$. In all cases, the fit quality is poor ($\chi^2_{red}>4$), indicating a 
consistent scatter in the normalized temperature profiles.
Introducing a dependence on $\sigma$ (Equation~\ref{fitvik06sigmaz}, with $\zeta=0$) clearly improves the fit (F-test probability $\sim99.96\%$).
The best-fit values indicate that the external part of the profiles is steeper for increasing values of $\sigma$, i.e. transitioning
from CC to NCC clusters. 
If we also take into account a dependence on the redshift of the temperature profiles (by fitting Equation~\ref{fitvik06sigmaz} and
leaving both $\eta$ and $\zeta$ free to vary),
we obtain the following relation that can be considered to be the closest possible to a universal law for temperature profiles of clusters at 
$0.1<z<0.9$ that can be attained with the present data
\begin{equation}
\label{universallaw}
\left\{\begin{array}{l}
\displaystyle \frac{T}{T_{500}}(r, \sigma, z)=1.14\!\pm\!0.05\frac{(x/0.045)^{\alpha}+\xi}{(x/0.045)^{\alpha}+1}\cdot\frac{1}{(1+(x/0.4)^2)^{\beta}},\\
\\
\displaystyle   \xi = 0.33\pm0.06 \cdot \phi_+(\sigma,\zeta=0),\\
\\
\displaystyle   \alpha = 3.85\!\pm\!1.79 \cdot \phi_-(\sigma,z),\\
\\
\displaystyle   \beta = 0.16\!\pm\!0.04 \cdot \phi_+(\sigma,z),\\
\\
\displaystyle   \phi_\pm(\sigma,z)\! =\! 1+0.23\!\pm\!0.11\cdot(1+\sigma)+0.17\!\pm\!0.09\cdot(1+z)\:. \end{array} \right.
\end{equation}

This relation shows that the steepening of the profiles with increasing $\sigma$ is still robust, and the behavior with the redshift agrees
with the results of \S~\ref{compint}, i.e., higher $z$ corresponds to steeper profiles.

\section{Conclusions}

We analyzed an {\em XMM-Newton} sample of 12 bright ($L_X>4\times10^{44}$ erg s$^{-1}$) galaxy clusters in the redshift range 
$0.4<z<0.9$ with an average temperature $kT>4.5$~keV. This sample was extracted from the {\em XMM-Newton} sample analyzed in BAL12
by selecting all clusters with at least 3000 {\tt MOS} net counts to obtain radial temperature profiles with a sizeable number of radial 
bins and reasonable errors on the temperature ($\sigma_{kT}/kT<15\%$).

Taking advantage of EPIC {\em XMM-Newton}'s high throughput and effective area, which makes it an ideal instrument for performing a spatially resolved
spectral analysis, this paper aimed at a systematic study of the temperature profiles in galaxy clusters at $z>0.4$, which is not currently provided
in the literature.
The results we obtained can be summarized as follows:

\begin{itemize}

\item We extracted temperature profiles for the 12 clusters in our sample. The cluster extension ranged from 
$\sim500$~kpc to $\sim1.3$~Mpc from the center. All  profiles were found to be declining toward larger radii.\\

\item The temperature profiles of the galaxy clusters in our sample, normalized by the mean temperature $T_{500}$, were found to be generally 
self-similar and could be well described by a function obtained by adapting the relation of \citet{vikhlinin06b} derived for lower redshift 
clusters.\\

\item We divided the sample into five CC and seven NCC clusters by introducing a pseudo-entropy ratio $\sigma$ and defining
a threshold $\sigma=0.6$ between CC and NCC clusters with the latter having $\sigma\ge0.6$. The profiles of the two subsamples were found to 
be different mainly in the inner regions, with the inner slope parameter assuming a positive value in the CC clusters ($\alpha=1.14\pm0.59$), 
to fit the temperature drop in the center, and a value consistent with zero in the NCC clusters ($\alpha=0.18\pm0.66$), to fit their
flat central profile. The large errors on the measurement of the external slope $\beta_0$ 
gave inconclusive results on the differences between the samples. Fitting the external regions ($r>0.15r_{500}$) with a simpler function
yielded no significant difference between the slopes of CC and NCC clusters ($\beta_0=0.39\pm0.12$ and $\beta_0=0.37\pm0.06$, respectively).
The lack of any significant difference between the two samples could be attributed to the small sample size, and therefore to the few
data points available for the fit, especially for the CC clusters.\\

\item We introduced a function of both $r$ and $\sigma$ to fit the data points of CC and NCC clusters together. In this case, the improved statistics 
allowed us to detect a significant dependence of the temperature profiles on the pseudo-entropy ratio $\sigma$, showing an indication that the outer part 
of the profiles becomes steeper for higher values of $\sigma$ (i.e. transitioning toward the NCC clusters). 
This behavior would agree with the results obtained in lower redshift galaxy cluster samples \citep[e.g.][]{baldi07} and with
the universal pressure profile derived by \citet{arnaud10} in galaxy clusters in the REXCESS local sample ($z<0.2$), 
which show a low dispersion, especially in the external regions.\\

\item In all attempted fits, no evidence of redshift evolution could be found within the redshift range sampled by our clusters ($0.4<z<0.9$).\\

\item We compared our sample with the intermediate cluster sample of \citet{leccardi08a} at $0.1<z<0.3$, finding significant differences in both the CC and
the NCC cluster samples. In particular, we found that CC clusters at $z>0.4$ shows a temperature dip in the center less deep with respect to
CC clusters at $z<0.3$, while NCC clusters at $z>0.4$ showed steeper temperature profiles with respect to NCC clusters at $z<0.3$. 
This can be caused by the circumstance that higher $z$ clusters are at a less advanced stage of their formation and did not have enough time to create a large massive hot
gas halo comparable in size and mass with that of lower redshift clusters.\\

\item We defined for the first time the closest possible relation to a universal law for the temperature profiles of galaxy clusters at $0.1<z<0.9$. This
relation shows a dependence on the state of relaxation of the clusters and the redshift.

\end{itemize}

Although some of the results obtained in this paper can in principle be biased by possible selection effects introduced in extracting the
sample from the XMM archive (\S~\ref{sampsel}), we stress that this is the first time that a systematic study of the temperature profiles in galaxy clusters 
at $z>0.4$ has been attempted. Additional deep XMM-Newton and Chandra observations, and most likely a new generation of X-ray observatories, 
would be needed to improve the current knowledge of the temperature distribution in the hot gas of galaxy clusters at high redshift.


\begin{acknowledgements}
We acknowledge financial contribution from contracts ASI-INAF I/023/05/0 and I/088/06/0.
FG acknowledges financial support from contract ASI-INAF I/009/10/0.
\end{acknowledgements}

\bibliographystyle{aa}
\bibliography{alesbib.bib}

\appendix
\section{Notes on individual clusters}\label{individual}

\noindent
{\it A851} -- A851, the lowest redshift cluster in the sample ($z=0.407$), has been observed by EPIC {\em XMM-Newton} for 
$\sim$56~ks (ObsID: 0106460101), of which $\sim41$~ks were useful for scientific analysis after removing the periods
of high flares. Its temperature profile is quite flat in the central regions with a temperature 
of $\sim6$~keV, showing a significant decline only in the outer radial bin 
($r>700$~kpc). As indicated by a value of $\sigma=0.896_{-0.179}^{+0.227}$, this cluster is classified as NCC.\\

\noindent
{\it RXCJ0856.1+3756} -- RXCJ0856.1+3756 ($z=0.411$) has been observed by XMM for $\sim29$~ksec (ObsID: 0302581801),
with $\sim24$~ks available for our analysis after cleaning the event files from soft proton flares.
Although the temperature profile consists of only four radial bins, a hint of a temperature drop in the central bin
($r<170$~kpc) is observed, although its value of $\sigma=0.723_{-0.172}^{+0.235}$ puts this cluster in the NCC category. 
The temperature profile then declines from $kT\approx9$~keV
down to $kT\approx5.5$~keV) in the outer two bins ($r>250$~kpc).\\

\noindent
{\it RXJ2228.6+2037} -- EPIC {\em XMM-Newton} observed the cluster RXJ2228.6+2037 ($z=0.412$) for a total of $\sim26$~ksec
(ObsID: 0147890101), almost entirely useful for our analysis. Although a hint of a drop in the temperature is observed in the 
central two radial bins ($r<150$~kpc), with
$kT\approx8$~keV in the central region, this cluster is classified as NCC ($\sigma=0.633_{-0.092}^{+0.100}$). 
The temperatures reaches its maximum ($kT\approx9$~keV) at $r\sim200$~kpc and then
declines smoothly down to $\sim4$~keV at radii larger than $\sim850$~kpc.\\

\noindent
{\it RXCJ1206.2-0848} -- The cluster RXCJ1206.2-0848 (z=0.440) has been observed twice by EPIC {\em XMM-Newton} (ObsIds:
0302581901, 0502430401), but one of the two observation (ObsID: 0302581901) was heavily affected by soft proton 
flares and was therefore discarded. The $\sim30$~ks of the other observation were almost entirely available for scientific
analysis. The temperature profile can be considered approximately constant around $kT\approx10$~keV, although several 
fluctuations are observed in the profile and the temperature shows a drop of $\sim3.5$~keV between the center and the 
adjacent radial bin. This cluster is classified as CC, presenting a value of $\sigma=0.413_{-0.055}^{+0.056}$.\\

\noindent
{\it IRAS09104+4109} -- IRAS09104+4109 (z=0.442) has been observed by EPIC {\em XMM-Newton} for a total of $\sim14$~ks ($\sim12$~ks
of clean exposure time, ObsID: 0147671001). The temperature shows a hint of a temperature drop in the center ($\Delta kT\sim0.5$~keV), 
and decreases from $\sim$6.5~keV to $\sim$4.5~keV between the region located at $\sim$100~kpc 
from the center and the one located at $\sim$200~kpc, remaining constant around the latter value going toward larger 
radii. This cluster shows a value of $\sigma=0.582_{-0.126}^{+0.143}$ and is therefore classified as NCC.\\

\noindent
{\it RXJ1347.5-1145} -- RXJ1347.5-1145 (z=0.451), representing the brightest cluster in our sample with its X-ray 
luminosity of $4.2\times10^{45}$ erg s$^{-1}$ \citep[0.1-2.4~keV band,][]{ebeling10}, has been observed for $\sim38$~ks 
($\sim32$~ksec after cleaning for flares) by {\em XMM-Newton} (ObsID: 0112960101). 
This cluster shows evidence of a very strong cool core with the temperature dropping from
$kT\approx13.5$~keV to $kT\approx10$~keV in the central bin ($r<90$~kpc) and is classified as a CC cluster
($\sigma=0.349_{-0.032}^{+0.037}$, the lowest value among the clusters in our sample). Although
some irregularities in the profile are present, the temperature then declines smoothly toward the outer regions,
reaching a value around $\sim6$~keV at $r>750$~kpc.\\

\noindent
{\it CLJ0030+2618} -- The cluster CLJ0030+2618 (z=0.500) has been observed three times by {\em XMM-Newton} (ObsIds: 0302581101,
0402750201, 0402750601) for a total of $\sim94$~ksec, of which $\sim70$~ksec were available after removing the high-flaring
intervals. The number of counts available did not allow us to divide the emission into more than three radial bins, with
the temperature profiles being approximately constant around $\sim5$~keV. CLJ0030+2618 is classified as a NCC cluster, as 
indicated by a $\sigma=0.766_{-0.170}^{+0.207}$.\\

\noindent
{\it MS0015.9+1609} -- MS0015.9+1609 (z=0.541) has been observed twice by XMM (ObsIds: 0111000101, 0111000201) for a total
of $\sim45$~ks ($\sim36$~ks of clean exposure time). The temperature profile declines smoothly from $kT\approx12$~keV 
measured at the center down to a minimum of $\sim7$~keV observed as far as $\sim900$~kpc from the center, making of it
a clear example of a NCC cluster ($\sigma=0.877_{-0.131}^{+0.173}$).\\

\noindent
{\it MS0451.6-0305} -- The galaxy cluster MS0451.6-0305 (z=0.550) has been observed by {\em XMM-Newton} (ObsId: 0205670101)
for a total of $\sim45$~ks. The observation was heavily affected by soft proton flares and we were able to use only
$\sim26$~ks of the original exposure.
The observed temperature shows a drop in the central 200~kpc ($kT\sim9$~keV) and is then quite constant around 
$\sim11$~keV between 200 and 450~kpc. It then declines to $\sim6$~keV at $r>450$~kpc. The value of $\sigma=0.555_{-0.087}^{+0.101}$
measured for this cluster puts it in the CC category.\\

\noindent
{\it MACSJ0647.7+7015} -- EPIC {\em XMM-Newton} observed twice (ObsIDs: 0551850401, 0551851301) the galaxy cluster 
MACSJ0647.7+7015 (z=0.591) for a total of $\sim145$~ks ($\sim88$~keV of clean exposure time).
The temperature observed in this cluster shows a smooth decline from the center ($kT\approx12$~keV) toward the
outskirts ($kT\approx6-7$~keV at $r>450~kpc$), making of it another obvious example of NCC cluster 
($\sigma=0.700_{-0.079}^{+0.091}$).\\

\noindent
{\it MACSJ0744.9+3927} -- Two EPIC observations of the cluster of galaxy MACSJ0744.9+3927 (z=0.698) are present in
the {\em XMM-Newton} archive (ObsIDs: 0551850101, 0551851201) accounting for a total exposure time of $\sim154$~ks ($\sim106$~ks
after cleaning for high-flaring periods). MACSJ0744.9+3927 is another examples of strong CC present
in our sample (as indicated also by $\sigma=0.451_{-0.048}^{+0.053}$), showing a temperature drop from $\sim10$~keV 
(at $\sim150$~kpc from the center) to $\sim7.5$~keV in the 
central radial bin ($r<100$~kpc). The temperature smoothly declines toward the outskirts and stabilizes around a
value of $\sim6$~keV at $r>500$~kpc.\\

\noindent
{\it CLJ1226.9+3332} -- CLJ1226.9+3332, the farthest cluster in our sample (z=0.890), has been observed twice by
EPIC {\em XMM-Newton} (ObsIds: 0070340501, 0200340101) for a total of $\sim139$~ks. Both observation were heavily affected
by soft proton flares (in particular $\sim75$\% of ObsID 0070340501 was affected by soft protons), leaving just 
$\sim77$~ks of clean exposure time available for scientific analysis. The temperature shows a sharp decline from the center
($kT\approx13$~keV) toward the outer radial bin as far as $\sim500$~kpc from the core ($kT\approx6$~keV). This cluster is
classified as a NCC ($\sigma=0.687_{-0.143}^{+0.183}$).\\


\end{document}